\documentclass[aps,pre,twocolumn,groupedaddress,longbibliography]{revtex4-1}
\usepackage{graphicx,color,subfigure,amsmath,floatrow}
\begin{document}

\title{Two-stage athermal solidification of semiflexible polymers and fibers}
\author{Joseph D. Dietz}
\author{Robert S. Hoy}
%\email{rshoy@usf.edu}
\affiliation{Department of Physics, University of South Florida, Tampa, FL, 33620}
\date{\today}

\begin{abstract}
We study how solidification of model freely rotating polymers under athermal quasistatic compression varies with their bond angle $\theta_0$.
All systems undergo two discrete, first-order-like transitions:\ entanglement at $\phi = \phi_E(\theta_0)$ followed by jamming at $\phi = \phi_J(\theta_0) \simeq (4/3 \pm 1/10)\phi_E(\theta_0)$.  
For $\phi < \phi_E(\theta_0)$, systems are in a ``gas'' phase wherein all chains remain free to translate and reorient.
For $\phi_E(\theta_0) \leq \phi \leq \phi_J(\theta_0)$, systems are in a liquid-like phase wherein chains are entangled.
In this phase, chains' rigid-body-like motion is blocked, yet they can still locally relax via dihedral rotations, and hence energy and pressure remain extremely small.
The ability of dihedral relaxation mechanisms to accommodate further compression becomes exhausted, and systems rigidify, at $\phi_J(\theta_0)$. 
At and slightly above $\phi_J$,  the bulk moduli  increase linearly with the pressure $P$ rather than jumping discontinuously, indicating these systems solidify via rigidity percolation. The character of the energy and pressure increases above $\phi_J(\theta_0)$ can be characterized via chains' effective aspect ratio $\alpha_{\rm eff}$.
Large-$\alpha_{\rm eff}$ (small-$\theta_0$) systems' jamming is bending-dominated and is similar to that observed in systems composed of straight fibers. 
Small-$\alpha_{\rm eff}$ (large-$\theta_0$) systems' jamming is dominated by the degree to which individual chains' dihedrals can collapse into compact, tetrahedron-like structures.
For intermediate $\theta_0$, chains remain in highly disordered globule-like configurations throughout the compression process; jamming occurs when entangled globules can no longer even locally relax away from one another.
\end{abstract}
\maketitle

\section{Introduction}

Jamming of semiflexible polymers and fibers is of broad scientific interest for several reasons. 
A key parameter for these systems is  the aspect ratio $\alpha$, which is commonly defined as the square root of the average ratio of the maximum to minimum eigenvalues of chains' radius of gyration tensors. 
As chains' stiffness increases, their configurations interpolate continuously between random walks with $\alpha \simeq 3.4$ \cite{rudnick87} to rigid rods with $\alpha \sim N$ (where $N$ is their degree of polymerization).
The synthetic macromolecular polymers in typical commodity plastics lie near the flexible end of this spectrum; their rigidity plays only a secondary role in their rheology and glassy-state mechanics.\cite{rubinstein03,roth16}
At the opposite end of the spectrum, carbon nanotubes and stiff biopolymers such as $F$-actin are rodlike in the absence of thermally induced bending.\cite{hinner98,picu11}

Many systems lie between these two limits.
The natural elastic fibers birds use to build their nests are rather stiff but not particularly straight, and hence have $\alpha$ that are significantly below their effective $N$.\cite{weiner20}.
Many biological structures, e.g.\ the collagen that gives our skin and tendons their elasticity, are semiflexible-fiber networks.\cite{Broedersz:2014aa}
Synthetic semiflexible elastic fibers have long been used in a wide variety of commodities such as textiles and steel wool,\cite{picu11} and are now attracting significant interest for their potential use in metamaterials.\cite{weiner20}

The tools of theoretical polymer physics and granular physics should be applicable to these systems because they are both polymer-like and (at most) weakly thermalized, yet they have attracted little interest from the soft matter physics community until very recently.
Athough jamming of rodlike grains has been fairly well studied,\cite{philipse96,williams03,desmond06,marschall18,langston15} theoretical analysis of the jamming of semiflexible-fiber-like grains with internal degrees of freedom and $1 \ll \alpha \ll N$ has only just begun.\cite{hoy17,weiner20}
Thus there is an opportunity to gain key insights into these systems through simulations of simple coarse-grained models that capture their essential features.

\begin{figure}[h]
\includegraphics[width=3.375in]{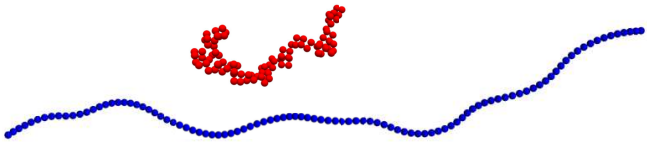}
\caption{Typical chain configurations at $\phi = \phi_J^{\rm mon}/3 \simeq .21$ for our highest and lowest aspect ratio polymers, i.e.\ FR polymers with $\theta_0 = 6^\circ$ (blue) and $\theta_0 = 90^\circ$ (red).  The apparent overlap of monomers in the $\theta_0 = 6^\circ$ chain is an illusion arising from the planar projection.}
\label{fig:twot0}
\end{figure}

Model freely rotating (FR) polymers are composed of $N$ tangent spheres of diameter $\sigma$, with fixed bond lengths ($\ell=\sigma$) and bond angles ($\theta=\theta_0$); see Figure \ref{fig:twot0}.
Unlike freely jointed (FJ) polymers, FR-polymer systems necessarily possess extensive frozen-in 3-body structural correlations arising from the fixed bond angles.
The distance between second-nearest intrachain neighbors is always $d_{13}(\theta_0) = 2\sigma\cos(\theta_0/2)$, where $\sigma$ is monomer diameter.
This $\theta_0$-dependent constraint significantly influences the structure of FR polymers' jammed states, even for the minimal $N = 3$.\cite{griffith19}
For larger $N$, these constraints causest FR polymers to exhibit jamming phenomenology \cite{hoy17} that is profoundly different than that of their FJ-polymeric counterparts.\cite{karayiannis08,karayiannis09b,karayiannis09c,reichhardt11,foteinopoulou08}
Whereas FJ polymers jam at $\phi \simeq \phi_J^{\rm mon} \simeq .64$, are isostatic at jamming, and possess polytetrahedral structural order very similar to that of jammed monomers,\cite{karayiannis08,karayiannis09b,karayiannis09c} FR polymers jam at significantly lower $\phi = \phi_J(\theta_0,N)$, their jammed states are quite hypostatic, and both their intrachain and interchain structural order at the $2$-, $3$-, and $4$-monomer levels depend strongly on $\theta_0$.\cite{griffith19,hoy17}

In this paper, we extend our previous studies\cite{hoy17,griffith19} by studying athermal solidification of long FR polymers in much greater detail.
We find that athermal solidification of these systems occurs in two distinct stages, characterized by two critical packing fractions:\ $\phi_E(\theta_0) $ and $\phi_J(\theta_0)$. 
At $\phi = \phi_E(\theta_0)$, systems undergo an entanglement transition where the system transitions from a disordered gas-like phase to a disordered liquid-like phase.
The entanglement transition is marked by sharp, first-order-like jumps in the average cooordination numbers for both monomers and chains.
However, chains can still locally relax via dihedral rotations, and hence systems' energy and pressure remain extremely small.
At $\phi = \phi_J(\theta_0) \simeq (4/3 \pm 1/10)\phi_E(\theta_0)$, systems jam (rigidity-percolate).
The cooordination numbers for both monomers and chains jump again, and the bulk modulus $B$ begins increasing linearly with the pressure $P$.
This two-stage process is a critical distinction between $\theta_0 > 0$ FR polymers and stiff fibers such as rods, spherocylinders, and $\theta_0 = 0$ FR-polymers, all of which jam when they entangle (i.e.\ have $\phi_E = \phi_J$ \cite{philipse96,williams03,rodney05}).

In addition to these universal features, systems exhibit several qualitative $\theta_0$-dependent differences.
Large-$\alpha$ (small $\theta_0$; see Figure \ref{fig:twot0}) systems' jamming is bending-dominated and is similar to that observed in systems composed of straight fibers.\cite{broedersz11,Broedersz:2014aa}
Chains tend to form large-scale arcs as systems are compressed; jamming occurs when these arcs can no longer bend further without an energy cost.
Small-$\alpha$ (large $\theta_0$) systems' jamming is dominated by the ability of chains to locally collapse into compact, tetrahedron-like structures.
Four-monomer chain segments become increasingly compact throughout the gas and liquid phases; jamming occurs when they can no longer collapse further without an energy cost.
For intermediate $\theta_0$, chains remain in highly disordered globule-like configurations throughout the compression.
These globules become more compact as compression proceeds. but do not locally collapse; jamming occurs when entangled globules can no longer even locally relax away from one another.

The rest of our paper is organized as follows.
In Section \ref{sec:methods}, we describe our FR-polymer model and the molecular-dynamics/energy-minimization algorithms we use to simulate athermal solidification of these systems.
In Section \ref{sec:results}, we describe our results for these systems' athermal solidification mechanisms and the mechanics of their jammed states in detail.
Finally, in Section \ref{sec:conclude} we summarize our results, discuss their potential implications for real systems, and conclude.

\section{Model and Methods}
\label{sec:methods}

\subsection{Freely rotating polymer model}

All systems are composed of $N_{ch} = 1600$ chains, each containing $N = 100$ monomers of mass $m$.  
All monomers interact via a harmonic potential $U_{H}(r) = 5\varepsilon (1 - r/\sigma)^2 \Theta(\sigma-r)$, where $\varepsilon$ is the energy scale of the pair interactions, $\sigma$ is monomer diameter, and $\Theta$ is the Heaviside step function.
Covalent bonds are modeled using the harmonic potential $U_c(\ell) = (k_c/2)(\ell-\sigma)^2$, leading to tangent-sphere polymers with equilibrium bond length $\ell_0=\sigma$.  
Angular interactions between three consecutive monomers along a chain are modeled by the harmonic potential $U_a(\theta) = (k_a/2)(\theta-\theta_0)^2$, where $\theta$ is the angle between consecutive bonds and is zero for straight trimers.
To capture a wide range of chain aspect ratios, we study systems with $\theta_0 = 3i^\circ$ for $i = 2, 3, ..., 30$.
We do not report data for systems with $\theta_0 < 6^\circ$ here because such systems are effectively in the elastic-rod-like ($\theta_0 = 0$) limit previously studied by Rodney, Picu and collaborators.\cite{rodney05,barbier09,subramanian11,picu11b}

Ideal FR chains are obtained in the limit $(k_c,k_a) \to \infty$.
The harmonic bond and angle potentials employed here limit the maximum MD timestep to $dt_{\rm max} \sim k_c^{-1}$.
To make our simulations computationally feasible, we choose $k_c = 300\varepsilon/\sigma^2$ and $k_a = 600\varepsilon/\rm{radians}^2$.
These parameter choices limit deviations from $\ell = \sigma$ to less than $10^{-3}\sigma$ and deviations from $\theta=\theta_0$ to less than $2^{\circ}$ under the conditions of primary interest here ($T = 0$ and $\phi <\sim \phi_J + .1$).
We will contrast results for these systems to those for fully-flexible ($k_a = 0$) chains, i.e.\ FJ chains.

\subsection{Sample preparation and compression protocol}

Studies of particulate granular systems can be conducted in a limit where $\phi_J$ is well defined \cite{torquato00} by employing initial states with packing fractions $\phi_{init}$ that we verified are low enough for $\phi_J$ to be $\phi_{init}$-independent.\cite{chaudhuri10}  
This is not the case for studies where the grains have internal degrees of freedom and can contract during compression.  
For FR polymers, the low-$\phi_{init}$ limit would produce chains that are maximally collapsed before they come into contact. 
Since this is not the type of experiment we wish to model, we choose to employ initial states with a common $\phi_{init} = \exp(-1.5)\phi_{cp} = .1652$, where $\phi_{cp} = \pi/\sqrt{18} \simeq .7405$ is the 3D close-packing density.
We emphasize that smaller or larger $\phi_{init}$ can produce different results.

We prepare our systems using standard molecular dynamics techniques. 
All MD simulations are performed using LAMMPS.\cite{plimpton95}
Initial states are generated by placing all chains with random positions and orientations within cubic cells of side length $L_0$, such that $\phi_{init} = \pi N_{ch}N/6L_0^3$.
Periodic boundary conditions are applied in all three directions.
Newton's equations of motion are integrated with a timestep $d t = .005\tau$, where the unit of time is $\tau=\sqrt{m\sigma^2/\varepsilon}$.
All systems are equilibrated at $T = \varepsilon/k_B$ until both intrachain and interchain structure have converged, then cooled to $T=0$ at a rate $10^{-5}\varepsilon/k_B\tau$.
For this value of $\phi_{init}$, all systems remain homogeneous and isotropic throughout this process, i.e.\ they do not develop nematic order.

After cooling, we simulate athermal solidification under quasistatic compression using alternating intervals of slow dynamic compression and energy minimization.
During dynamic compression, $L$ is varied in time as $L(t) = L_0 \exp(-\dot\epsilon t)$.  
We choose $\dot\epsilon = 10^{-6}/\tau$, which is the slowest rate feasible for our employed system size ($N_{ch}N= 1.6\cdot10^5$).
At increments $\Delta \phi/\phi = -.001$ (i.e.\ every time density has increased by $0.1\%$), we stop this compression and allow systems to relax and minimize their energy.
We find that the optimal energy-minimization strategy for our systems is letting them relax via damped Newtonian dynamics, i.e.\ using the equation of motion $m\ddot{\vec{r}}_i = \vec{F}_i  - \dot{\vec{r}}_i/\tau_{\rm damp}$ for all monomers, where $\vec{F}_i$ is the usual Newton's-2nd-law force on monomer $i$ and $\tau_{\rm damp} = 10\tau$ is the damping time.
This dynamics allows monomers to push completely off one another during energy minimization, and yields lower total system energies $E_{\rm tot}(\phi)$ than the Polak-Ribi{\'e}re congugate-gradient,\cite{polak69} Hessian-free truncated-Newton,\cite{grippo89}  or FIRE \cite{bitzek06} algorithms.
We find that the minimal minimization time per cycle that yields converged results for our $N = 100$ systems is $\tilde{\tau}_{min} \simeq 4\cdot10^4\tau$; all results presented below are for this $\tilde{\tau}_{min}$.
After energy minimization, we restart the dynamic compression and repeat this compression-minimization cycle until $\phi= .67$.

Jamming is defined to occur when the pressure $P$ within energy-minimized states exceeds $P_{thres}= 10^{-6}\varepsilon/\sigma^3$.
As in our previous study,\cite{hoy17} we choose to identify jamming with the emergence of a finite bulk modulus \cite{vanHecke09} rather than with the vanishing of soft modes \cite{ohern03} because proper handling of soft modes associated with ``flippers'' (interior monomers with zero or one noncovalent contacts \cite{karayiannis09b}) is highly nontrivial.

\subsection{Measuring polymers' $\phi$-dependent aspect ratios}

An essential difference between rigid-rod-like particles (e.g.\ spherocylinders and large-$\alpha$ ellipsoids) and the systems considered here is that the latter have internal degrees of freedom (i.e.\ their $N-3$ dihedral angles) and can reconfigure during compression.  
Their aspect ratio $\alpha$ is not fixed; it decreases with increasing $\phi$ as chains adopt more compact configurations.
Specifically, it is given by
\begin{equation}
\alpha_{\rm eff}(\phi) = \bigg{\langle} \sqrt{ \displaystyle\frac{ A_i(\phi) } {\max[a_i(\phi), \sigma^2/6]} } \bigg{\rangle},
\label{eq:alphaeff}
\end{equation}
where $A_i(\phi)$ and $a_i(\phi)$ are respectively the maximum and minimum values of chain $i$'s radius of gyration tensor, and the average is taken over all chains.
Ideal random walks have $\alpha_{eff} = \alpha_{\rm RW} \simeq 3.435$.\cite{rudnick87}
To compare our results for FR polymers to analogous results for rods and ellipsoids,\cite{philipse96,williams03,donev04,donev07} we monitor how our systems' $\alpha_{\rm eff}$ vary during compression.

\begin{figure}[h]
\includegraphics[width=3.25in]{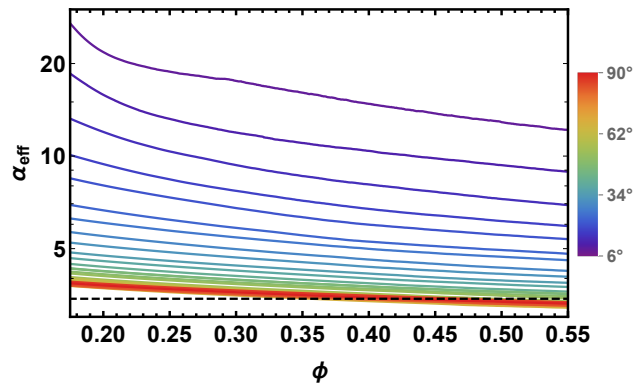} 
\caption{$\phi$-dependent aspect ratios for all systems.  The color-scale legend indicates $\theta_0$, and the dashed horizontal lines indicates $\alpha_{\rm RW}$.  
Lower $\phi_{init}$ lead to lower $\alpha_{\rm eff}$ for any given $\phi$ in regimes where $\alpha_{\rm eff} \gg \alpha_{\rm RW}$.
Note that $\alpha_{\rm RW}$ is the lower bound  of $\alpha_{\rm eff}$ only for random-walk-like configurations; spherelike collapsed globules have  $\alpha_{\rm eff} \simeq 1$.}
\label{fig:alphaphi}
\end{figure}

Figure \ref{fig:alphaphi} shows $\alpha_{\rm eff}(\phi)$ for all systems.
$\alpha_{\rm eff}$ decreases significantly with $\phi$ for all $\theta_0$ even when chains are not in contact [i.e.\ for $\phi \ll \phi_E(\theta_0$)] because the equations of motion employed during our dynamic compression intervals are those of standard strain-controlled molecular dynamics, i.e. 
\begin{equation}
\begin{array}{ccc}
\dot\vec{r}_i = \dot\vec{v}_i - \dot{\bar{\epsilon}}\cdot \vec{r}_i & \ \ \ \textrm{and} \ \ \ & \dot\vec{p}_i = \vec{F}_i +  \dot{\bar{\epsilon}}\cdot \vec{p}_i
\end{array},
\label{eq:straincontrol}
\end{equation}
where $\vec{r}_i$, $\vec{v}_i$ and $\vec{p}_i$ are respectively the position, velocity and momentum of monomer $i$, $\vec{F}_i$ is the total force on monomer $i$, and $\dot{\bar{\epsilon}}$ is the true strain rate tensor.\cite{hoover80}
This protocol mimics embedding the polymers in a medium (e.g.\ a dense solvent) that favors affine contraction under hydrostatic compression.  
It becomes equivalent to the standard MD protocol for simulating jamming of particulate systems, where \textit{instantaneous} finite increments $\Delta \phi/\phi$ are imposed,\cite{ohern03} in the limit $\dot\epsilon \to \infty$.
However, it is distinct from a ``solventless'' compression that lacks the $\dot{\bar{\epsilon}}\cdot \vec{r}_i$ terms, i.e.\ it is distinct from a protocol that shrinks the simulation cell without moving monomers.
Such a protocol would reduce the rate of decrease of $\alpha_{\rm eff}(\phi)$, particularly for $\phi < \phi_E(\theta_0)$.
Other details of the results shown above will be discussed below.

\section{Results}
\label{sec:results}

In this Section, we present results for systems' jamming densities, hypostaticities, stress-strain curves and how they break down into pair vs.\ bond vs.\ angular contributions, liquid (entangled) phase structure, microscopic (sub-chain-scale) jamming mechanisms, and  stress transmission mechanisms vary with $\theta_0$.
All results presented below are averages over three independently prepared systems.
We emphasize that the below results are \textit{not} for the long-chain (large-$N$) limit, especially for $\theta_0 < 20^\circ$.\cite{hoy17}
This limit is reached only when $N \gg C_\infty$, where $C_\infty$ is the statistical segment length above which chains become random-walk-like.
Chains in this limit have $\alpha_{\rm eff} \simeq \alpha_{\rm RW}$, a property which is incompatible with the present effort to study \textit{semiflexible} chains with $1 \ll \alpha \ll N$.

\subsection{Densities and aspect ratios at jamming}

\begin{figure}[h]
\includegraphics[width=3.25in]{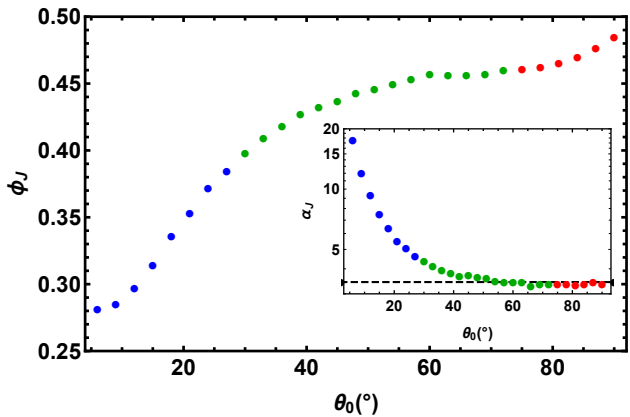} 
\caption{$\theta_0$-dependence of  $\phi_J$ and $\alpha_J$.  Blue, green, and red symbols respectively indicate the small-, intermediate-, and large-$\theta_0$ regimes. The horizontal dashed line in the inset indicates $\alpha = \alpha_{\rm RW}$.}
\label{fig:PJAJ}
\end{figure}

Figure \ref{fig:PJAJ} shows $\phi_J(\theta_0)$ for all systems.
As we previously reported,\cite{hoy17}  $\phi_J(\theta_0)$ increases monotonically with $\theta_0$.
The quasistatic protocol employed here enables large-scale stress-relaxation processes that are frozen out for dynamic compression for currently feasible $\dot{\epsilon}$; for this reason, the $\phi_J(\theta_0)$ shown here are larger than those reported in Ref.\ \cite{hoy17}.
For $\theta_0 <\sim 9^\circ$, systems are in a rigid-rod-like limit where $\phi_J$ is almost $\theta_0$-independent (albeit $\phi_{init}$-dependent).
Beyond this limit, $\phi_J$ increases rapidly, with an inflection point at $\theta_0 \simeq 15^\circ$.
The rate of increase $\partial \phi_J/\partial \theta_0$ decreases with increasing $\theta_0$, and $\phi_J$ nearly plateaus for $60^\circ <\sim \theta_0 < 75^\circ$.
Finally, another inflection point makes $\phi_J(\theta_0)$ concave up for $\theta_0 \geq 75^\circ$, and it continues to increase slowly all the way up to $\theta_0 = 90^\circ$.
We will show in the following sections that $\theta_0 < 27^\circ$, $27^\circ \leq \theta_0 < 75^\circ$, and $\theta_0 \geq 75^\circ$ systems define three regimes with qualitatively different jamming phenomenology.
For the remainder of the paper we will refer to these as the small-, intermediate-, and large-$\theta_0$ regimes, or alternatively as the high-, intermediate-, and low-aspect ratio regimes.

As shown in Fig.\ \ref{fig:alphaphi}, chains' $\alpha_{eff}(\phi)$ decreases smoothly with increasing $\phi$; no jumps are apparent either at $\phi_E(\theta_0)$ or $\phi_J(\theta_0)$.  
This indicates that chains' structure on scales comparable to their radii of gyration evolves continuously in all systems.
Thus it is worthwhile to measure their effective aspect ratio at jamming, $\alpha_J = \alpha_{\rm eff}(\phi_J)$.
Results for all systems are shown in the inset to Fig.\ \ref{fig:PJAJ}. 
$\theta_0 = 6^\circ$ polymers have $\alpha_J \simeq 17$, which is comparable to yet significantly below $\alpha_{\rm RR} \simeq 28$ (the $\alpha_{\rm eff}$ value for ideally rigid $N = 100$ $\theta_0 = 0$ chains with $a_i = \sigma^2/6$).
These chains are well into the semiflexible regime defined by $\alpha_{\rm RW} < \alpha_J < \alpha_{\rm RR}$.
$\alpha_J(\theta_0)$ decreases monotonically with increasing $\theta_0$ before plateauing at $\alpha_J \simeq \alpha_{\rm RW}$ for $\theta_0 \geq 60^\circ$.
$N = 100$ FR polymers with $\theta_0 \geq 60^\circ$ are therefore random-walk-like when they jam.  
We will show below that this random-walk-like structure leads the marginally jammed states to share some common features, but that other features remain qualitatively different, especially for low- vs.\ moderate-aspect-ratio chains.

\subsection{Hypostaticity of marginally jammed states}

\begin{figure}[h]
\includegraphics[width=3.25in]{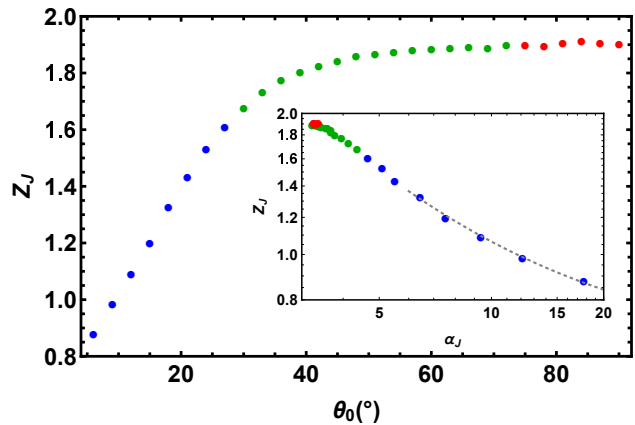} 
\caption{$\theta_0$- and $\alpha_J$-dependencies of  $Z_J$.  The dotted gray curve in the inset shows $Z_J = .61674 + 4.48568/\alpha$.}
\label{fig:ZJ}
\end{figure}

A freely rotating $N$-mer has $n_{\rm dof} = N + 3$ degrees of freedom:\ 3 rigid translations, 3 rigid rotations, and $N-3$ dihedral angles.
The Maxwell criterion for jamming suggests that these systems should jam at $Z_{J} = Z_{\rm iso} \equiv 2n_{\rm dof}/N = 2 + 6/N$.
On the other hand, aspherical particles ranging from rods \cite{philipse96} to superballs \cite{jiao10} always jam hypostatically.
Moreover, we previously showed that FR polymers jam hypostatically under dynamic compression, with a $Z_J$ that decreases with decreasing $\theta_0$ and with increasing $N$.\cite{hoy17}
Here we examine their hypostaticity in more detail.
Figure \ref{fig:ZJ} shows $Z_J(\theta_0)$ and $Z_J(\alpha_J)$ for the current systems; only noncovalent contacts are included in $Z_J$.
Note that the $Z_J(\theta_0)$ shown here are significantly larger than those of Ref.\ \cite{hoy17} because they are for quasistatic rather than dynamic compression; here the pressure defining $\phi_J$ comes from a larger number of smaller-overlap contacts.

As in our previous study,\cite{hoy17} $Z_J$ increases smoothly and monotononically with $\theta_0$ before plateauing for $\theta_0 > 60^\circ$.
This large-$\theta_0$ plateau in marginally jammed systems' hypostaticity $H(\theta_0) = 2 + 6/N - Z_J(\theta_0)$ is of fundamental interest.
One expects jammed FR polymers to be at least somewhat hypostatic since the forces and torques transmitted along chains via their bond and angular interactions do not contribute to $Z_J$.
Previous results showing FJ polymers jam isostatically \cite{karayiannis08,karayiannis09b,karayiannis09c} were obtained for tangent hard-sphere models that lack bond tensions.
Another potential source of the hypostaticity is that although our model does not include friction explicitly, friction is nonetheless present.\footnote[1]{The frictional and frictionless isostaticity criteria, $Z_J = 2n_{dof}/N$ and $Z_J = n_{dof}/N + 1$, are almost identical for our systems because $n_{dof}/N = 1 + 3/N \simeq 1$.}
The friction arises from chains' pearl-necklace structure, and can be understood in terms of their local concavity and their tendency to interlock.\cite{schreck09,ludewig12} 
Our previous work on bent-core trimers  \cite{griffith19} suggests that the present systems' $\phi_J(\theta_0)$ would be increased and their $H(\theta_0)$ reduced if chains had greater monomer overlap, i.e.\ equilibrium bond lengths $\ell_0 < \sigma$.

We find that $Z_J$ decreases smoothly and monotononically with increasing $\alpha_J$.
Results for our highest-aspect-ratio systems are consistent with $Z_J \sim \alpha_J^{-1}$, which is consistent with results for rigid rods and $\theta_0 = 0$ FR polymers.\cite{philipse96,rodney05}
Comparing the results shown in Fig.\ \ref{fig:ZJ} to those shown in Fig.\ \ref{fig:PJAJ} makes it evident that $Z_J$ is primarily controlled by $\alpha_J$ rather than $\phi_J$.
More generally, the data show some of the trends observed for rods \cite{philipse96}, $\theta_0 = 0$ FR polymers,\cite{rodney05,barbier09} and ellipsoids,\cite{donev04} but also major qualitative differences.

Axisymmetric rigid rods' coordination number at jamming becomes $\alpha$-independent for $\alpha > 15$, while their density continues to decrease as $\phi_J \sim \alpha^{-1}$.\cite{philipse96}
Rodlike ($\theta_0 = 0$) FR polymers show similar behavior; as few as 4 total interchain contacts are sufficient to produce jamming.\cite{rodney05}
Clearly early even our stiffest systems, which have $N Z_J(6^\circ) \simeq 88$ interchain contacts at jamming, are very far from this limit.
This is unsurprising given that $\theta_0 > 0$ FR polymers possess internal degrees of freedom, but the large variation of $Z_J$ with $\theta_0$ and $\alpha_J$ shown in Fig.\ \ref{fig:ZJ} demonstrates that semiflexible-polymer jamming is controlled by fundamentally different physics than that of both their sitff and flexible counterparts.

Ellipsoids jam hypostatically because their spatially varying local curvature can block rotational motions.\cite{donev07}
Their degree of hypostaticity $H(\alpha) = 2n_{\rm dof} - Z_J(\alpha)$ decreases with increasing $\alpha$, and saturates at its minimal value $H \simeq 0.1$ for $\alpha > \alpha^*$, where $\alpha^*$ is the aspect ratio that maximizes $\phi_J$.\cite{donev04,donev07}
FR polymers with $\theta_0 >\sim 60^\circ$ have $\alpha_J \simeq \alpha_{\rm RW}$ and $H(\theta_0) \simeq .17$; this value is comparable to but slightly larger than ellipsoids' minimal $H(\alpha)$.
However, since FR polymers' $H(\theta_0)$ increases with increasing chain length,\cite{hoy17} this good agreement may be coincidental.

\subsection{Mechanics of jammed systems}

For $\phi < \phi_E(\theta_0)$, all systems' pressures and potential energies remain at zero to within the accuracy of our minimization algorithm.
For $\phi_E(\theta_0) < \phi < \phi_J(\theta_0)$, pressures become finite, but remain extremely small.
The onset of solidification is marked by continuous increases in the bulk moduli $B(\phi) = \phi \frac{dP}{d\phi}$ from zero.
The initial stages of these increases have $B \propto P$, indicating that  $\theta_0 > 0$ FR polymers, like their $\theta_0 = 0$ counterparts, \cite{rodney05,barbier09,subramanian11,picu11b} solidify via rigidity percolation.\cite{thorpe85}
Jamming of semiflexible polymers and fibers must therefore be regarded as a \textit{continuous} transition.

\begin{figure}[h]
\includegraphics[width=3.25in]{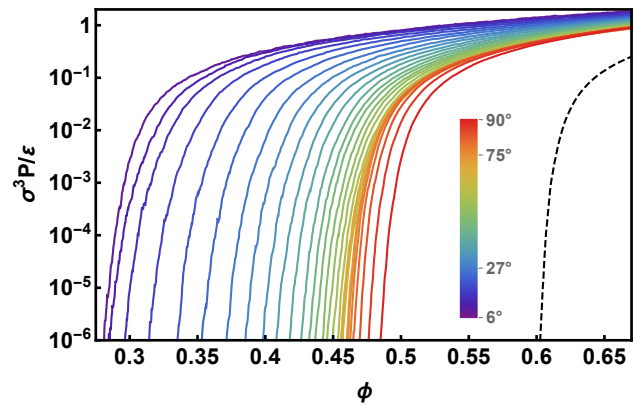} 
\caption{Pressure $P(\phi)$ for all FR polymers (colored curves) compared to $P(\phi)$ for FJ polymers (black dashed curve).  The legend indicates $\theta_0$.}
\label{fig:ss1}
\end{figure}

As discussed above (Sec.\ \ref{sec:methods}), we defined $\phi_J$ using the criterion $P_{thres} = P(\phi_J) =  10^{-6}\varepsilon/\sigma^3$. 
Figure \ref{fig:ss1} shows all systems' stress-strain curves for $\phi > \phi_J(\theta_0)$.
While $B(\phi_J)$ depends rather strongly on $\theta_0$, all systems' $B(\phi_J)$ are approximately proportional to their $\phi_J$;  $B(\phi_J)/\phi_J$ is nearly constant for $\theta_0 \leq 60^\circ$, then increases by a factor of less than two as $\theta_0$ increases from $60^\circ$ to $90^\circ$.
The scalings of $B$ with $(\phi - \phi_J)$ are similar to those previously reported for $\theta_0 = 0$ systems.\cite{rodney05,barbier09}

The small-but-noticeable jumps in the $P(\phi)$ curves correspond to stress-relieving plastic avalanches that are much like those discussed in Ref.\ \cite{subramanian11}; these jumps are more prominent when results for single samples rather than averages over independently prepared samples are plotted.
Such plastic avalanches are more frequent and begin at lower pressures in the small-$\theta_0$ (large-$\alpha_{\rm eff}$) systems; this is expected since they are less dense and chains have more room to rearrange at fixed $(\phi - \phi_J)$.

\begin{figure}[h]
\includegraphics[width=3.25in]{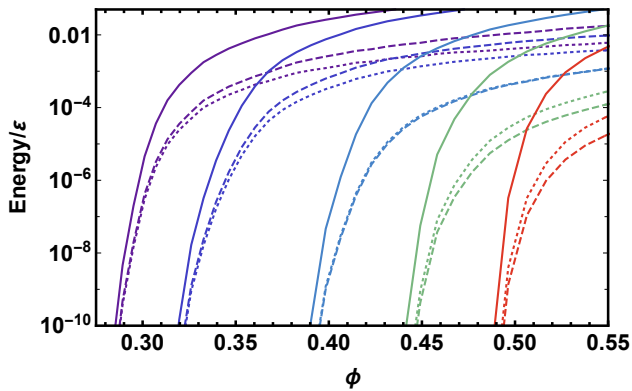}
\caption{
Energies as a function of system density for FR polymers with (from left to right) $\theta_0 = 6^\circ$, $15^\circ$, $27^\circ$, $45^\circ$, and $90^\circ$. Solid, dotted and dashed curves respectively show $\varepsilon_{\rm pair}(\phi)$, $\varepsilon_{\rm bond}(\phi)$, and $\varepsilon_{\rm angle}(\phi)$.  For $\theta_0 = 27^\circ$, $\varepsilon_{\rm bond}(\phi)$ and $\varepsilon_{\rm angle}(\phi)$ are nearly equal and the curves overlap.}
\label{fig:ss2}
\end{figure}

More insight can be gained by breaking down the mechanical response into contributions from pair, bonded, and angular terms.
Since the contribution of 3-body angular interactions to any system's virial (and hence its pressure) is identically zero,\cite{bekker94} we analyze them in terms of their associated energies rather than directly in terms of the stress-strain curves.
Figure \ref{fig:ss2} shows how selected systems'  per-monomer total, pair, bond, and angle energies $\varepsilon_{\rm tot}(\phi)$, $\varepsilon_{\rm pair}(\phi)$, $\varepsilon_{\rm bond}(\phi)$, and $\varepsilon_{\rm angle}(\phi)$ vary with $\phi$.
All systems have $\varepsilon_{\rm pair}(\phi) > \varepsilon_{\rm bond}(\phi)$ and  $\varepsilon_{\rm pair}(\phi) > \varepsilon_{\rm angle}(\phi)$; this is expected since intermonomer overlap is the ultimate origin of both entanglement and jamming.

The variation of $\varepsilon_{\rm angle}(\phi)/\varepsilon_{\rm bond}(\phi)$, however, is far less trivial.
For our $N$ and preparation protocol,  $\varepsilon_{\rm angle}(\phi) > \varepsilon_{\rm bond}(\phi)$ for $\theta_0 \leq 27^\circ$.
This bending-dominated jamming\cite{broedersz11,rodney05} defines our small-$\theta_0$/large-$\alpha_J$ regime.
In these systems, jamming is caused by chains' resistance to large-scale bending, and specifically by the torques transmitted along entangled chain segments.
For $\theta_0 > 27^\circ$, $ \varepsilon_{\rm bond}(\phi) > \varepsilon_{\rm angle}(\phi)$.
Thus both our intermediate-$\theta_0$ and large-$\theta_0$ systems exhibit stretching-dominated jamming, where the dominant factor promoting jamming is chains' resistance to axial compression.
We find that all systems' per-monomer $\varepsilon_{\rm tot}$, $\varepsilon_{\rm pair}$, $\varepsilon_{\rm bond}$, and $\varepsilon_{\rm angle}$ each scale approximately as $P^2$ immediately above $\phi_J$.
These scalings are the same as they are for rodlike $\theta_0 = 0$ FR polymers.\cite{rodney05}
However, as illustrated in Fig.\ \ref{fig:ss2}, the associated prefactors in the scalings $\varepsilon_{\rm tot} \sim \tilde{t}(\theta_0)P^2$, $\varepsilon_{\rm pair} \sim \tilde{p}(\theta_0)P^2$, $\varepsilon_{\rm bond} \sim \tilde{b}(\theta_0)P^2$, and $\varepsilon_{\rm angle} \sim \tilde{a}(\theta_0)P^2$ vary strongly with $\theta_0$.
The ratios $\tilde{a}(\theta)/\tilde{t}(\theta_0)$, $\tilde{b}(\theta)/\tilde{t}(\theta_0)$, and $\tilde{a}(\theta)/\tilde{b}(\theta_0)$ all increase monotonically with decreasing $\theta_0$.

We emphasize that the results presented above are sensitive to the ratio $k_a/k_c$.
Broederz \textit{et al.}\cite{broedersz11} showed that varying the ratio of angular and axial stiffnesses over a wide range can qualitatively change the nature of systems' mechanics, e.g.\ from bending-dominated to stretch-bend coupled to stretching-dominated as $k_a/k_c$ increases.
This dependence is complicated and nonmonotonic; for example, stretching dominates bending both for $k_a = 0$ and in the $k_a/k_c \to \infty$ limit.
Limited computational resources prevent us from exploring this issue in greater detail here, but it would be an interesting topic for future studies.

\subsection{Jamming's precursor: Entanglement}

\begin{figure}[h]
\includegraphics[width=3.25in]{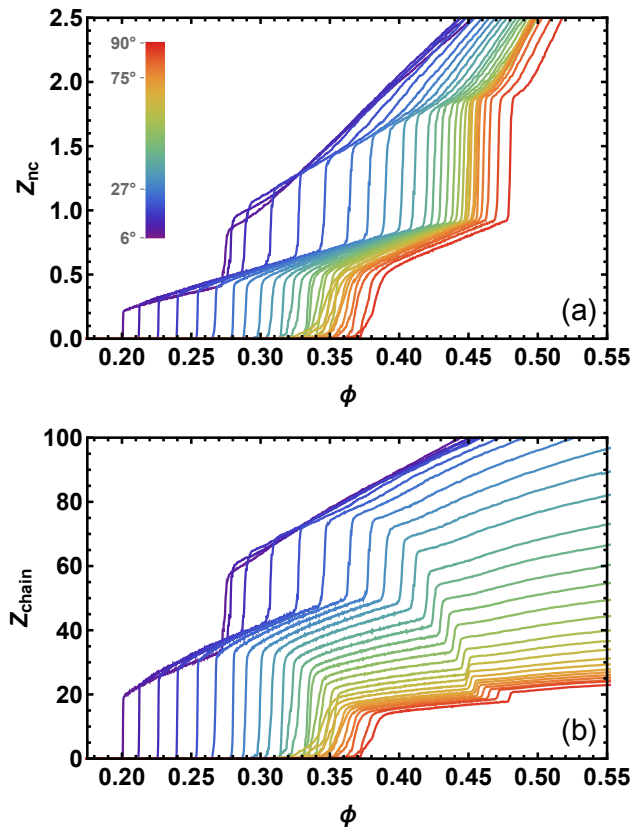} 
\caption{$\phi$-dependence of chain and monomer coordination.  Panel (a):\ $Z_{\rm chain}(\phi)$, the average number of other chains contacted by a given chain.  Panel (b):\ $Z_{nc}(\phi)$, the average number of noncovalent contacts per monomer.  Data for flexible chains follows similar trends:\ entanglement at $\phi_E \simeq .515$ followed by jamming at $\phi_J \simeq .60$.}
\label{fig:Zjumps}
\end{figure}

As mentioned above, we find that FR polymers' athermal solidification under quasistatic compression occurs in two distinct stages:\ entanglement at $\phi = \phi_E(\theta_0)$ followed by jamming at $\phi = \phi_J(\theta_0) \simeq (4/3 \pm 1/10)\phi_E(\theta_0)$.
Figure \ref{fig:Zjumps} shows that these transitions can be clearly identified in terms of the average coordination numbers for monomers and chains.
$Z_{nc}(\phi)$ is the coordination number discussed in the preceding sections, i.e.\ the average number of noncovalent contacts per monomer.
$Z_{\rm chain}(\phi)$ is the average number of other chains that contact a given chain.
For all $\theta_0$, both stages of athermal solidification are marked by  first-order-like jumps in both $Z_{\rm chain}(\phi)$ and $Z_{nc}(\phi)$.

The nearly-linear increases in $Z_{nc}(\phi)$ and $Z_{\rm chain}$ for $\phi_E < \phi < \phi_J$ and $\phi > \phi_J$ can be simply explained using a binary-contact model.
If one assumes the probability of interchain contacts can be understood at the two-chain level (i.e.\ higher-order correlations between chain configurations are unimportant), the total number of contacts should scale as $\phi^2$, and thus the number of contacts per chain should scale as $\phi$.
This assumption is equivalent to assuming that an entangled segment corresponds to a fixed number of interchain contacts, and correctly predicts the entanglement densities of concentrated polymer solutions.\cite{degennes74,colby92}
The first-order-transition-like jumps of $Z_{nc}(\phi)$ and $Z_{\rm chain}(\phi)$ indicate that the assumption breaks down at $\phi_E(\theta_0)$ and $\phi_J(\theta_0)$.

Below $\phi_E(\theta_0)$, $Z_{nc}(\phi)$ and $Z_{\rm chain}(\phi)$  are zero because chains are free to relax away from one another via rigid-body-like rotations and translations with no energy cost. 
These relaxation mechanisms become increasingly less viable as density increases.  
Previous studies of large-$\alpha$ particulate systems have shown that if no locally nematic alignment occurs, the available particle rotations become severely blocked as $\phi$ increases, and entanglement occurs when they vanish.\cite{williams03,desmond06}
A similar process occurs in our systems, with larger-$\alpha_{\rm eff}$ chains entangling at lower $\phi$.
Entanglement at $\phi = \phi_E$ is a contact percolation transition;\cite{picu11} percolation of the interchain contact network is what prevents chains from fully relaxing away from each other as they did for $\phi < \phi_E$.
Such contact percolation is an intrinsically $\sim N_{ch}$-body phenomenon. 
As discussed above, FR polymers' jamming at $\phi_J$ is a rigidity-percolation-like transition.
When a percolating network of load-bearing contacts is formed, the system's energy can no longer relax to zero.
This is an intrinsically $\sim N_{ch}N$-body phenomenon.
The many-body nature of contact and rigidity percolation explains why the binary-contact model breaks down at $\phi_E$ and $\phi_J$.\footnote[2]{This two-stage mechanism appears to be specific to \textit{quasistatic} compression; FR polymers under dynamic compression exhibit increases in $Z_{\rm chain}(\phi)$ that are gradual rather than sharp, and viscous ($\dot\epsilon$-dependent) stresses within the liquid phase.\cite{hoy17}}

The $\theta_0$-dependence of the height of the jumps in $Z_{nc}$ at $\phi = \phi_E$ appear to indicate that binary-contact scaling ``switches on'' at $\phi = \phi_E$.
For $\phi_E(\theta_0) < \phi \leq \phi_J(\theta_0)$, the data nearly collapse onto a common line $Z_{nc}(\phi) \simeq 3\phi - .4$, indicating that the three-body $\theta = \theta_0$ constraints play a minimal role in this regime.
These ``switching events'' are nearly instantaneous for our lowest-$\theta_0$ systems.
They become more gradual as $\theta_0$ increases, but remain present even for our largest-$\theta_0$ systems.
The height $\Delta Z_{\rm chain}(\theta_0)$ of the jumps in $Z_{\rm chain}$ at $\phi = \phi_E$ is also $\theta_0$-dependent, but in a manner
that can be qualitatively understood as a tradeoff between $\phi$-dependent and $\alpha_{\rm eff}$-dependent effects.
For $\theta_0 <\sim 24^\circ$, $\Delta Z_{\rm chain}$ increases with $\theta_0$ because the total number of monomer-monomer contacts formed at entanglement increases with $\phi_E$.
For $\theta_0 >\sim 24^\circ$, $\Delta Z_{\rm chain}$ decreases with $\theta_0$.
In this regime, the total number of monomer-monomer contacts formed at entanglement remains nearly $\theta_0$-independent over a broad range of $\theta_0$, but since chains become less spatially extended as  $\alpha_{\rm eff}$ decreases, these contacts are distributed between fewer distinct pairs of chains, and thus $\Delta Z_c$ becomes smaller.

\begin{figure}[h]
\includegraphics[width=3.25in]{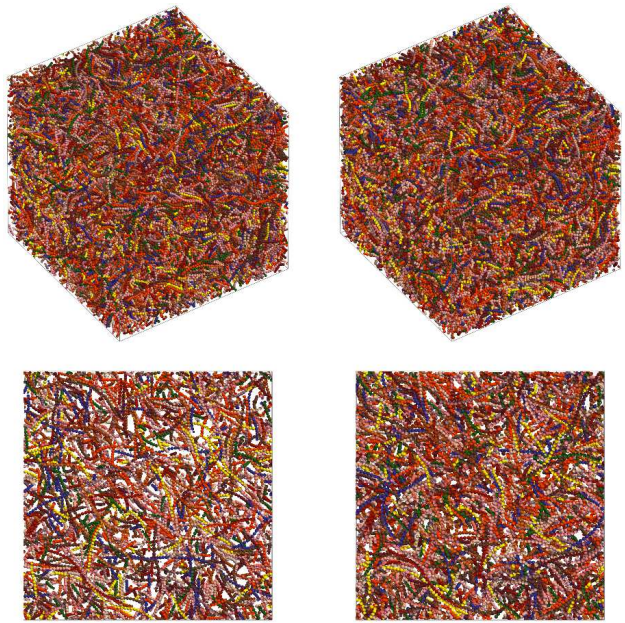}
\caption{Snapshots of $\theta_0 = 6^\circ$ FR-polymer configurations at $\phi_e$ (left panels) and $\phi_J$ (right panels).  The top panels show all chains, while the bottom panels show cross-sections of thickness $4.25\sigma$.  Different colors indicate different chains.}
\label{fig:snapshots}
\end{figure}

Figure \ref{fig:snapshots} shows snapshots of our $\theta_0 = 6^\circ$ systems at $\phi_E$ and $\phi_J$.  
At $\phi_E$, it is clear that chains are entangled and cannot move in a rigid-body-like fashion, but it is also clear that many interior chain segments remain free. 
At $\phi_J$, there are clearly far fewer of these free interior segments, and hence fewer available relaxation mechanisms.
These qualitative observations can be made quantitative by plotting the flipper fraction $F_{\rm flip}(\phi)$, i.e.\ the probability that interior monomers have zero interchain contacts.\footnote[3]{Our previous study \cite{hoy17} defined flippers as interior monomers with less than two noncovalent contacts.  We find that the more restrictive definition employed here clarifies the relevant physics.}
Each flipper corresponds to an unconstrained dihedral DOF.
Flippers do not prevent systems from being mechanically stable; forces and torques can be transmitted through flippers along chain backbones.
An example of a mechanically stable material with a high flipper fraction is rubber, which is clearly a solid despite the fact that its constituent chains fluctuate freely on the microscopic, sub-crosslink scale.

\begin{figure}[h]
\includegraphics[width=3.25in]{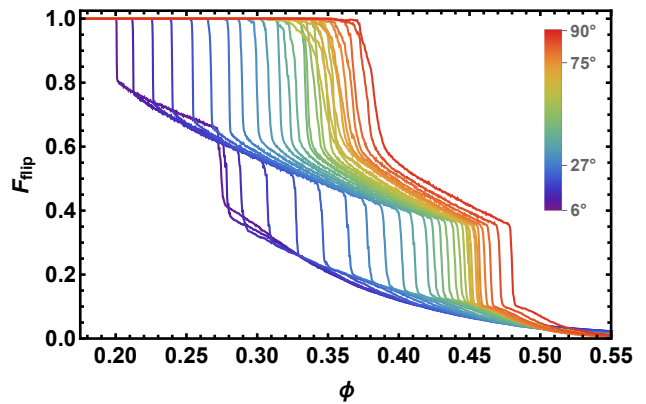}
\caption{$\phi$-dependence of the flipper fraction:\ $F_{\rm flip}(\phi)$ is the fraction of flippers (interior monomers with zero noncovalent contacts).}
\label{fig:flips}
\end{figure}

Results for all systems' $F_{\rm flip}(\phi)$ are shown in Figure \ref{fig:flips}.
For $\phi < \phi_E$, all systems have $F_{\rm flip} = 1$, as expected.
 $F_{\rm flip}$ drops sharply at both $\phi_E$ and $\phi_J$; these drops' qualitative trends with $\theta_0$ reflect the trends in $Z_{nc}$ discussed above.
All systems retain a substantial fraction of flippers even above $\phi_J$, indicating that (as expected) these systems' mechanical integrity is maintained by forces and torques transmitted along chain backbones.
More noteworthy, however, is the behavior for $\phi_E(\theta_0) < \phi < \phi_J(\theta_0)$.
All systems' $F_{\rm flip}(\phi)$ drop approximately as $\phi^{-1}$, which indicates that chains' interior segments are continuously becoming more constrained as compression proceeds.
This process corresponds to chains rearranging (with negligible energy cost) via dihedral rotations.

\subsection{Mechanisms of chain collapse}

As discussed above, for $\phi_E \leq \phi \leq \phi_J$, systems are in a liquid-like phase wherein chains cannot rearrange on large scales due to entanglement, yet they can still fully relax via dihedral rotations.
Here we will show that examining the evolution of systems' dihedral-angle distributions between $\phi_E$ and  $\phi_J$ reveals critical differences between our small-, intermediate, and large-$\theta_0$ systems' solidification mechanisms, and also clarifies the reasons for their different mechanics above $\phi_J$.

\begin{figure}[h]
\includegraphics[width=3.25in]{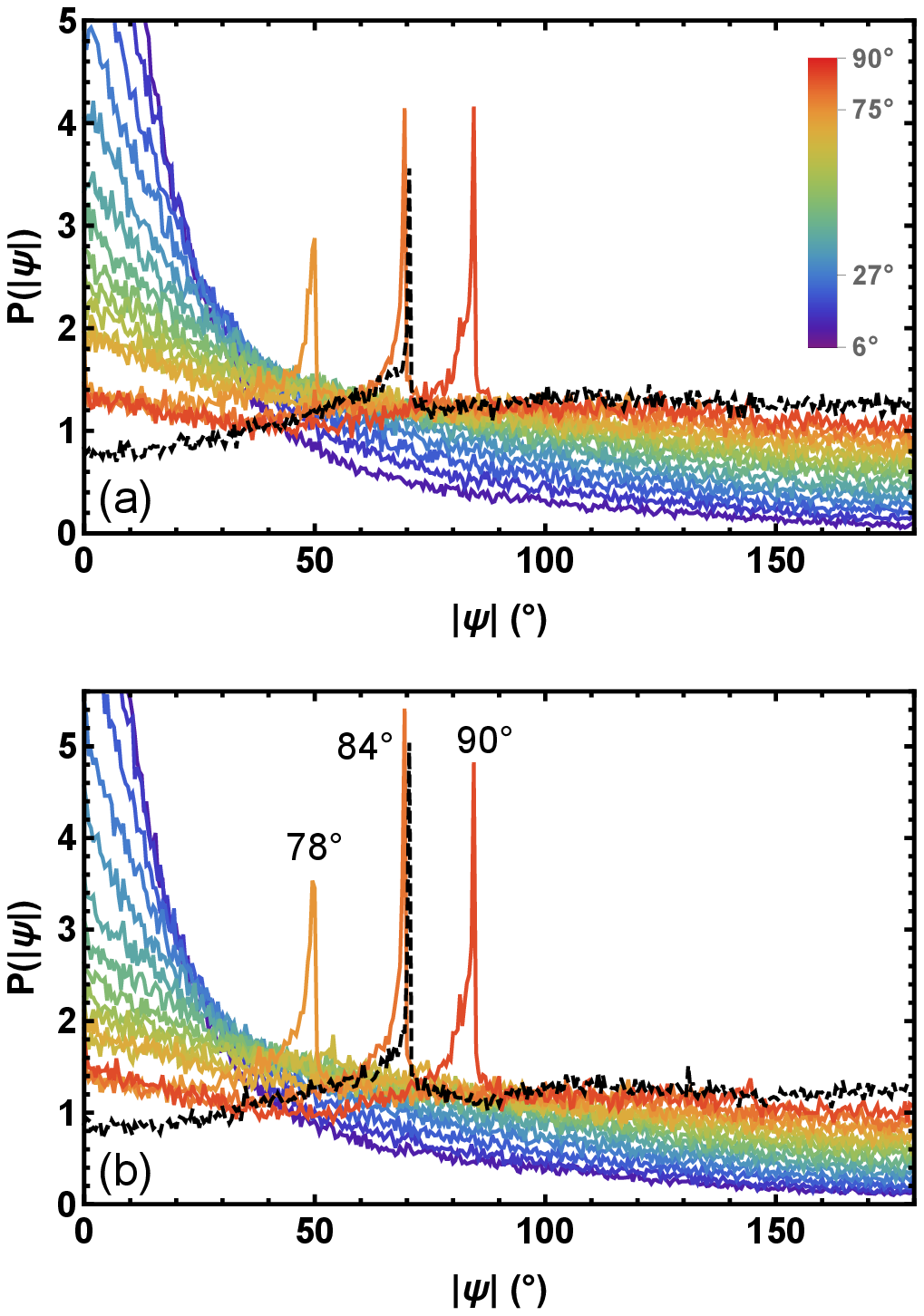} 
\vspace{-15pt}
\includegraphics[width=3.25in]{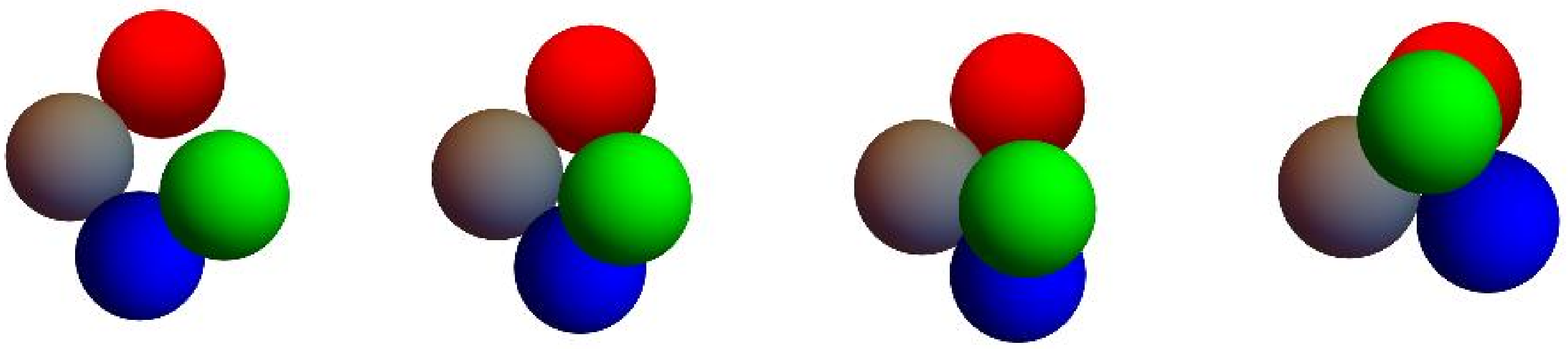}
\caption{Dihedral angle distributions at $\phi_E$ (panel a) and $\phi_J$ (panel b).  These are normalized so that totally disordered systems have $P(\Psi) = 1$.  The black dashed curve indicates results for FJ chains. For clarity, results are shown only for every other $\theta_0$, i.e.\ $\theta_0 = 6^\circ$, $12^\circ$, ..., $90^\circ$.  The angle labels in panel (b) indicate which $\theta_0$ the respective peaks correspond to.  The bottom panel shows snapshots of the most probable dihedral arrangement for $\theta_0 = 78^\circ$, $84^\circ$, $90^\circ$, and flexible chains, from left to right.}
\label{fig:dihedEJ}
\end{figure}

The  dihedral angle $\psi$ is the angle between the planes defined by two consecutive trimers along a chain.
Specifically, if $\vec{b}_{ij} = \vec{r}_j - \vec{r}_i$ is the covalent bond vector connecting monomers $i$ and $j $, and $\{ i, j, k, l \} = \{ i, i+1, i+2, i+3 \}$ are four consecutive monomers along a given chain:
\begin{equation}
|\Psi| = cos^{-1} \left( \displaystyle\frac{(\vec{b}_{ij} \times \vec{b}_{jk})\cdot (\vec{b}_{jk} \times \vec{b}_{kl})}{| \vec{b}_{ij} \times \vec{b}_{jk} | | \vec{b}_{jk} \times \vec{b}_{kl}|} \right).
\label{eq:psi}
\end{equation}
Figure \ref{fig:dihedEJ} shows all systems' dihedral-angle distributions $P(|\psi|)$ at $\phi_E$ and $\phi_J$.
Here $P(\psi)$ is normalized so that completely disordered systems have $P(\psi) = 1$; our systems satisfy this condition at $\phi = \phi_{init}$.
Thus $P(\psi) > 1$ ($P(\psi) < 1$) indicates that dihedrals with the angle $\Psi$ become more (less) likely during compression.

Small-$\theta_0$ systems develop an excess of \textit{cis} ($\Psi \simeq 0$) conformers even below $\phi_E$.
These excess \textit{cis} conformers tend to be grouped consecutively along chains, forming increasingly large-scale arcs like those shown in Fig.\ \ref{fig:twot0}.\footnote[4]{$n_\theta = 360^\circ/\theta_0$ consecutive $\Psi = 0$ conformers form a circle; smaller numbers of consecutive $\Psi = 0$ conformers form a planar circular arc.}
Between $\phi_E$ and $\phi_J$, entanglement prevents these large-scale circular arcs from disappearing or loosening.
Instead they slowly tighten, and jamming occurs when they lock.
Above $\phi_J$, further compression of these arcs requires work.
For $\theta_0 < 27^\circ$ this process dominates systems' mechanical properties as shown in Fig.\ \ref{fig:ss2}.

For intermediate $\theta_0$, $P(|\Psi|)$ remains relatively flat as compression proceeds.
A slight excess of \textit{cis} conformers is balanced by a slight deficit of \textit{trans} ($|\Psi \simeq 180^\circ|$) conformers.
This is consistent with a gradual collapse into disorded, globule-like structures that jam when chains are no longer able to collapse into more compact configurations without incurring an energy cost.

In contrast, large-$\theta_0$ systems' athermal solidification is dominated by \textit{local} collapse and ordering of chains at the few-monomer scale.
Previous studies have shown that flexible chains develop a peak at $|\Psi| = 70.53^\circ$ that corresponds to formation of locally polytetradral order.\cite{anikeenko07,karayiannis09b}
FR polymers with $\theta_0 \neq 120^\circ$ cannot form such tetrahedra due to their $\theta = \theta_0$ constraints.
However, their dihedrals \textit{do} tend to collapse into the maximally compact structures consistent with both the $\theta = \theta_0$ constraints and polytetrahedral-like order, as indicated by the different peaks in $P(|\Psi|)$.\footnote[5]{This collapse mechanism appears to be specific to \textit{quasistatic} compression; large-$\theta_0$ FR polymers under dynamic compression do not develop sharp peaks in $P(|\Psi|)$.\cite{hoy17}}
As compression continues, the number of these compact dihedrals increases as indicated by the increase in the height of these peaks.
Jamming occurs when the dihedral DOF available for further chain collapse without energy cost become exhausted.
As illustrated in the snapshots, the most likely dihedral arrangements become increasingly compact as $\theta_0$ increases.
This directly explains the increase in $\phi_J$ with $\theta_0$ for $\theta_0 >\sim 75^\circ$.

\subsection{Stress transmission in marginally jammed states}

We conclude our analyses by examining how the local stresses on individual monomers are distributed.
The Cauchy stress invariants $I_1$, $I_2$ and $I_3$ are defined in terms of the elements of the Cauchy stress tensor $\bar{\sigma}$ as 
\begin{equation}
\begin{array}{ccccc}
I_1 = \sigma_{ii}  & , & I_2 = \frac{1}{2} ( \sigma_{ii} \sigma_{jj} - \sigma_{ij} \sigma_{ji}) & , & I_3 = \textrm{det}( \bar{\sigma}) ,
\end{array}
\label{eq:cauchyinv}
\end{equation}
where repeated indices indicate summation over $\{ i, j, k \} = \{ 1, 2, 3 \}$.
These quantities remain well-defined when $\bar{\sigma} \equiv \bar{\sigma}^i$ is the \textit{atomic}-level stress tensor for atom $i$.
Since the macroscopic pressure $P$ is given by
\begin{equation}
P = -\displaystyle\frac{1}{3N_{atom}} \sum_{i = 1}^{N_{atom}} I_1^i,
\label{eq:Ptot}
\end{equation}
examining how the atomic-level $\{ I_1^i \}$ are distributed is useful for characterizing stress inhomogeneities in jammed systems.\cite{ohern01}

\begin{figure}[h]
\includegraphics[width=3.25in]{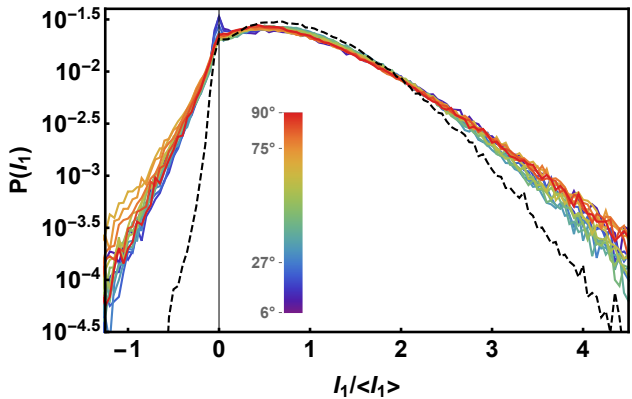} 
\caption{Probability distributions for the atomic-level principal stress $I_1$ at $\phi = \phi_J(\theta_0)$.  For clarity, results are shown only for every other $\theta_0$, i.e.\ $\theta_0 = 6^\circ$, $12^\circ$, ..., $90^\circ$.  The black dashed curve shows data for FJ chains.}
\label{fig:pofI1}
\end{figure}

The probability distributions $P(I_1)$ for all marginally jammed systems are shown in Figure \ref{fig:pofI1}.
All systems' $P(I_1)$ are maximal at $I_1 = 0$ due to their hypostaticity; note that the peak sharpens with decreasing $\theta_0$ as $\phi_J$ decreases and $H(\theta_0)$ increases.
The monomers with $I_1 = 0$ are either chain ends or flippers.
Another distinctive feature  is the very broad peak in $P(I_1)$ at $I_1/\langle I_1 \rangle \simeq 3/5$. 
Other marginally jammed systems, both particulate and polymeric,\cite{ohern01,rottler02b} tend to have $P(I_1)$ that decay nearly perfectly exponentially away from their peaks.
Since a comparably broad peak is not found in dense polymer glasses,\cite{rottler02b} the one observed here probably also results from our systems' hypostaticity and lower $\phi_J$.

Unlike model particulate granular systems, which typically have purely repulsive interactions and hence single-sided $P(I_1)$,\cite{ohern01,ohern03} polymers have a double-sided $P(I_1)$ that indicates some monomers in these states are under tension.
These tensile forces can only come from the covalent backbone bonds; angular forces do not directly contribute.\cite{bekker94}
Such double-sided distributions have been previously observed in fully developed glassy-polymer crazes.\cite{rottler02b}
The presence of significant tensile forces in these marginally jammed states is a major difference between granular polymers/fibers and their particulate counterparts such as spheres, rods, and ellipsoids.

Our data show that angular interactions significantly broaden FR polymers' $P(I_1)$ distributions relative to those found in FJ polymers for all $\theta_0$, but increasingly so as $\theta_0$ increases.
The greater stress inhomogeneities within FR polymers' marginally jammed states probably arises from an interplay between their lower $\phi_J$ and greater ability to transmit stress.
For $I_1 >\sim 2 \langle I_1 \rangle$, the $P(I_1)$ distributions depend rather strongly on $\theta_0$.
Higher-aspect-ratio chains' $P(I_1)$ decay faster, indicating these systems possess fewer highly stressed monomers, i.e.\ that the compressive stresses giving rise to systems' finite $P$ are less localized.
One plausible explanation for this observation is that the larger-$\alpha$ chains are better at transmitting forces along their backbones.

To test this hypothesis, we examined whether and how our systems' monomer-level stresses are topologically correlated.
The most compressive principle stress on a given monomer, $\sigma_3$, is given by the minimum eigenvalue of $\bar{\sigma}$ and is often used to analyze force-chain networks in particulate systems.\cite{peters05}
We calculated $\sigma_3$ for all monomers, and labeled monomers with above-average compressive stress (i.e.\ a monomer with $\sigma_3 < \langle \sigma_3 \rangle$) as ``overcompressed.''
Then we calculated $P(n)$, the probability that a monomer $i$ lying a chemical distance $n$ away from an overcompressed monomer $j$ is also overcompressed.
Topologically uncorrelated stresses would give $P(n) = 1/2$ for all $n > 0$.

\begin{figure}[h]
\includegraphics[width=3.25in]{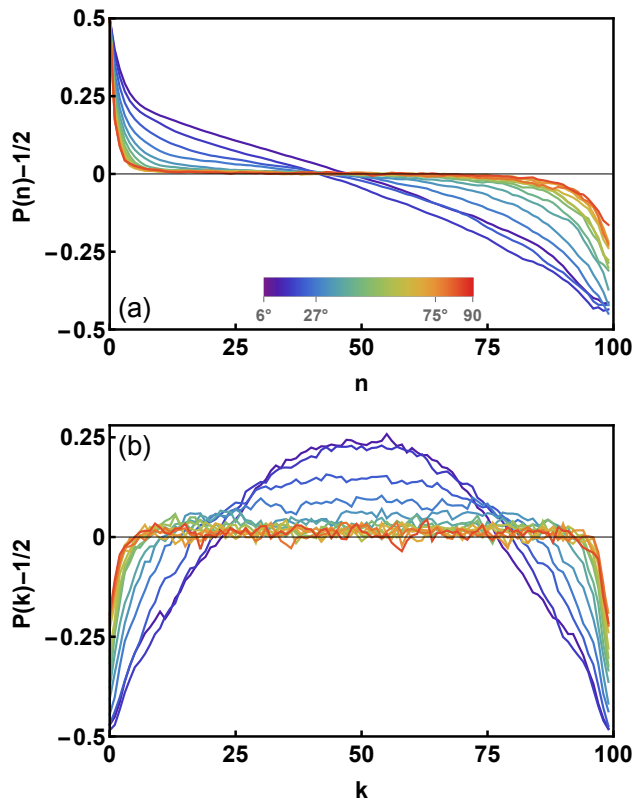}
\caption{Stress transmission along chain backbones and its relation to chain-end effects.  Panel (a):\ $P(n)$, the probability that a monomer $i$ lying a chemical distance $n$ away from an overcompressed monomer $j$ is also overcompressed (i.e.\ has $\sigma_{3}^i < \langle \sigma_3 \rangle$).  Panel (b):\ $P(k)$, the probability the $k^{th}$ monomer in a chain is overcompressed. }
\label{fig:topo}
\end{figure}

Results for $P(n)$ for all systems are shown in Figure \ref{fig:topo}(a).
For small- and intermediate-$\theta_0$ systems, these correlations are minimal.
FR polymers with  $\theta_0 >\sim 30^\circ$ exhibit slightly longer-range correlations than their FJ counterparts, but their $P(n)$ still decay to $1/2$ by $n \simeq 5$.
This suggests that these systems' slightly longer-range stress correlations arise from their restricted dihedral-level structure.
For small-$\theta_0$ systems with bending-dominated jamming, however, the correlations are significantly longer-ranged.
These systems' stress-transmission mechanisms can be better understood if one first examines the distribution of overcompressed monomers along chain backbones.

Fig.\ \ref{fig:topo}(b) shows $P(k)$, the probability that the $k$th monomer along a chain is overcompressed.
For intermediate- and large-$\theta_0$ systems, all monomers are approximately equally likely to be overcompressed except those very near to chains' free ends.
As $\theta_0$ decreases into the bending-dominated regime, however, overcompressed monomers become more and more concentrated towards the middle of chains.
The simplest explanation for this result is that the large-scale arcs that low-$\theta_0$ chains form as $\phi$ increases \cite{hoy17} are stabilized by overcompression of their interiors. 
This stretch-bend coupling may be the single mechanism most responsible for jamming in our small-$\theta_0$ systems.
Comparable stretch-bend-coupling-dominated mechanics are observed in a wide variety of fibrous materials.\cite{picu11,broedersz11,Broedersz:2014aa}

The increasing concentration of overcompressed monomers towards the center of chains directly explains the longer-range correlations in $P(n)$ for large-aspect-ratio polymers/fibers.
$P(n) - 1/2$ decays slower than exponentially because overcompressed sections of chains are more likely to be arranged consecutively as opposed to randomly along chains.
It also explains the anticorrelation in $P(n)$ for $n > 50$.
Since overcompressed monomers are less likely to appear near chain ends, it becomes especially unlikely that two monomers a chemical distance $n > N/2$ will both be overcompressed -- increasingly so as $\theta_0$ decreases.

\section{Summary and Conclusions}
\label{sec:conclude}

In this paper, we examined how model semiflexible polymers athermally solidify under quasistatic compression.
Our findings show a number of universal features, but also indicate that these systems' solidification mechanisms are highly sensitive to chains' local and large-scale structure, i.e.\ their bond angle $\theta_0$ and their effective aspect ratio $\alpha_{\rm eff}$. 
They also show that larger-aspect ratio polymers' athermal solidification exhibits many features found in previous studies of semiflexible \textit{fibers}.\cite{picu11,weiner20}

All of our systems soldify in two distinct, sharply defined and well-separated stages.
For $\phi < \phi_E(\theta_0)$, systems are in a gas-like phase wherein chains are able to avoid contact and can thus move in rigid-body-like fashion.
For $\phi_E(\theta_0) \leq \phi < \phi_J(\theta_0)$, systems are in an liquid-like phase wherein chain entanglement prevents rigid-body-like motion but chains can accommodate further compression via local rearrangements of their internal degrees of freedom, i.e.\ dihedral rotations.
In this phase, the entanglement density scales as $\phi^2$, as predicted by binary-contact models and found in concentrated polymer solutions.\cite{degennes74,colby92}
When the abovementioned dihedral relaxation mechanisms can no longer accormmodate further compression, systems rigidity-percolate and jam at $\phi_J(\theta_0)$. 
The large separation between the entanglement and jamming densities, $\phi_J(\theta_0)/\phi_E(\theta_0) = 4/3 \pm 1/10$, is a critical distinction between semiflexible polymers and their more rigid counterparts such as rods, ellipsoids, and $\theta_0 = 0$ FR polymers, all of which jam when they entangle.\cite{philipse96,williams03,donev04,rodney05}

Higher-aspect-ratio chains entangle and jam at lower densities because they are more spatially extended.
They are also far more hypostatic at jamming, presumably because they are better able to transmit forces and torques along their backbones.
One interpretation of these results is that higher-aspect-ratio chains' \textit{effective} number of DOF is well below $N-3$ and decreases with decreasing $\theta_0$; after all, axisymmetric rigid rods have $n_{dof} = 5$.
This interpretation is consistent with previous results for both thermal and athermal bent-core trimers, where the minimization of configurational freedom coincides with the minimization of $\phi_J$\cite{griffith19} and maximization of both the glass transition temperature $T_g$\cite{meena19} and the crystallization temperature $T_x$\cite{salcedo19} as $\theta_0 \to 0$.
It may be more useful to regard these chains as being composed of $N/n_{corr}$ segments, where $n_{corr}$ is the chemical length over which stresses are topologically correlated.
These segments may correspond to the large circular arcs (of $\sim n_{corr}$ consecutive \textit{cis}-conformers) they form under compression.  
Testing these ideas requires further study.

Higher-aspect-ratio chains also exhibit different mechanics above jamming than their lower-aspect ratio counterparts.
As $\alpha_J$ increases, systems' resistance to compression is increasingly dominated by resistance to large-scale chain bending.
This trend is analogous to that of elastic-rod-like fibers, whose mechanics become increasingly bending-dominated as their length increases.\cite{rodney05,picu11}
As aspect ratio decreases, bending is less important because large circular arcs no longer form.
Instead, systems' resistance to compression becomes increasingly dominated by their axial stiffness, i.e.\ the stiffness of their covalent bonds.
A comparable switch from bending-dominated to stretching-dominated jamming occurs in model lattice-based networks as their coordination number increases \cite{broedersz11} (as is the case in our systems; $Z_J$ increases with decreasing $\alpha_J$).
Finally, for polymers with $\theta_0 >\sim 75^\circ$, chains' dihedrals tend to collapse into the most compact structures consistent with both their $\theta = \theta_0$ constraints and locally polytetrahedral-like order, even below $\phi_J$.
These compact chain segments likely play a critical role in these systems' mechanics.

Here we have considered only homogeneous, monotonically compressed systems.
One natural extension of our work would be to examine inhomogeneous fibrous systems such as those found in bird nests, synthetic nonwoven materials, and novel metamaterials inspired by these.\cite{picu11,weiner20}
The results presented above suggest that these can be modeled using large-aspect-ratio (small-$\theta_0$) chains.
In contrast, lower-aspect-ratio (larger-$\theta_0$) chains are a useful model for colloidal and granular polymers, which are also inhomogeneously structured in typical experiments.\cite{zou09,brown12,vutukuri12,mcmullen18,dumont18} 
Another natural extension of our work would be to examine systems that were mechanically preconditioned by successive compression-decompression cycles.
While it is known that comparable systems exhibit substantial mechanical hysteresis,\cite{barbier09,subramanian11,picu11,weiner20} the degree to which this hysteresis varies with fiber aspect ratio or ``waviness'' remains largely unexplored.

This material is based upon work supported by the National Science Foundation under Grant DMR-1555242.

%\bibliography{polyjam,trimers,glassproposal,cites2,PPA3}

\begin{thebibliography}{58}%
\makeatletter
\providecommand \@ifxundefined [1]{%
 \@ifx{#1\undefined}
}%
\providecommand \@ifnum [1]{%
 \ifnum #1\expandafter \@firstoftwo
 \else \expandafter \@secondoftwo
 \fi
}%
\providecommand \@ifx [1]{%
 \ifx #1\expandafter \@firstoftwo
 \else \expandafter \@secondoftwo
 \fi
}%
\providecommand \natexlab [1]{#1}%
\providecommand \enquote  [1]{``#1''}%
\providecommand \bibnamefont  [1]{#1}%
\providecommand \bibfnamefont [1]{#1}%
\providecommand \citenamefont [1]{#1}%
\providecommand \href@noop [0]{\@secondoftwo}%
\providecommand \href [0]{\begingroup \@sanitize@url \@href}%
\providecommand \@href[1]{\@@startlink{#1}\@@href}%
\providecommand \@@href[1]{\endgroup#1\@@endlink}%
\providecommand \@sanitize@url [0]{\catcode `\\12\catcode `\$12\catcode
  `\&12\catcode `\#12\catcode `\^12\catcode `\_12\catcode `\%12\relax}%
\providecommand \@@startlink[1]{}%
\providecommand \@@endlink[0]{}%
\providecommand \url  [0]{\begingroup\@sanitize@url \@url }%
\providecommand \@url [1]{\endgroup\@href {#1}{\urlprefix }}%
\providecommand \urlprefix  [0]{URL }%
\providecommand \Eprint [0]{\href }%
\providecommand \doibase [0]{http://dx.doi.org/}%
\providecommand \selectlanguage [0]{\@gobble}%
\providecommand \bibinfo  [0]{\@secondoftwo}%
\providecommand \bibfield  [0]{\@secondoftwo}%
\providecommand \translation [1]{[#1]}%
\providecommand \BibitemOpen [0]{}%
\providecommand \bibitemStop [0]{}%
\providecommand \bibitemNoStop [0]{.\EOS\space}%
\providecommand \EOS [0]{\spacefactor3000\relax}%
\providecommand \BibitemShut  [1]{\csname bibitem#1\endcsname}%
\let\auto@bib@innerbib\@empty
%</preamble>
\bibitem [{\citenamefont {Rudnick}\ and\ \citenamefont
  {Gaspari}(1987)}]{rudnick87}%
  \BibitemOpen
  \bibfield  {author} {\bibinfo {author} {\bibfnamefont {J.}~\bibnamefont
  {Rudnick}}\ and\ \bibinfo {author} {\bibfnamefont {G.}~\bibnamefont
  {Gaspari}},\ }\bibfield  {title} {\enquote {\bibinfo {title} {The shapes of
  random walks},}\ }\href@noop {} {\bibfield  {journal} {\bibinfo  {journal}
  {Science}\ }\textbf {\bibinfo {volume} {237}},\ \bibinfo {pages} {384}
  (\bibinfo {year} {1987})}\BibitemShut {NoStop}%
\bibitem [{\citenamefont {Rubinstein}\ and\ \citenamefont
  {Colby}(2003)}]{rubinstein03}%
  \BibitemOpen
  \bibfield  {author} {\bibinfo {author} {\bibfnamefont {M.}~\bibnamefont
  {Rubinstein}}\ and\ \bibinfo {author} {\bibfnamefont {R.}~\bibnamefont
  {Colby}},\ }\href@noop {} {\emph {\bibinfo {title} {Polymer Physics}}}\
  (\bibinfo  {publisher} {Oxford University Press},\ \bibinfo {year}
  {2003})\BibitemShut {NoStop}%
\bibitem [{\citenamefont {Roth}(2016)}]{roth16}%
  \BibitemOpen
  \bibfield  {author} {\bibinfo {author} {\bibfnamefont {C.B.}\ \bibnamefont
  {Roth}},\ }\href@noop {} {\emph {\bibinfo {title} {Polymer Glasses}}}\
  (\bibinfo  {publisher} {CRC Press},\ \bibinfo {year} {2016})\BibitemShut
  {NoStop}%
\bibitem [{\citenamefont {Hinner}\ \emph {et~al.}(1998)\citenamefont {Hinner},
  \citenamefont {Tempel}, \citenamefont {Sackmann}, \citenamefont {Kroy},\ and\
  \citenamefont {Frey}}]{hinner98}%
  \BibitemOpen
  \bibfield  {author} {\bibinfo {author} {\bibfnamefont {B.}~\bibnamefont
  {Hinner}}, \bibinfo {author} {\bibfnamefont {M.}~\bibnamefont {Tempel}},
  \bibinfo {author} {\bibfnamefont {E.}~\bibnamefont {Sackmann}}, \bibinfo
  {author} {\bibfnamefont {K.}~\bibnamefont {Kroy}}, \ and\ \bibinfo {author}
  {\bibfnamefont {E.}~\bibnamefont {Frey}},\ }\bibfield  {title} {\enquote
  {\bibinfo {title} {Entanglement, elasticity, and viscous relaxation of actin
  solutions},}\ }\href@noop {} {\bibfield  {journal} {\bibinfo  {journal}
  {Phys. Rev. Lett.}\ }\textbf {\bibinfo {volume} {81}},\ \bibinfo {pages}
  {2614} (\bibinfo {year} {1998})}\BibitemShut {NoStop}%
\bibitem [{\citenamefont {Picu}(2011)}]{picu11}%
  \BibitemOpen
  \bibfield  {author} {\bibinfo {author} {\bibfnamefont {R.~C.}\ \bibnamefont
  {Picu}},\ }\bibfield  {title} {\enquote {\bibinfo {title} {Mechanics of
  random fiber networks?a review},}\ }\href@noop {} {\bibfield  {journal}
  {\bibinfo  {journal} {Soft Matt.}\ }\textbf {\bibinfo {volume} {7}},\
  \bibinfo {pages} {6768} (\bibinfo {year} {2011})}\BibitemShut {NoStop}%
\bibitem [{\citenamefont {Weiner}\ \emph {et~al.}(2020)\citenamefont {Weiner},
  \citenamefont {Bhosale}, \citenamefont {Gazzola},\ and\ \citenamefont
  {King}}]{weiner20}%
  \BibitemOpen
  \bibfield  {author} {\bibinfo {author} {\bibfnamefont {N.}~\bibnamefont
  {Weiner}}, \bibinfo {author} {\bibfnamefont {Y.}~\bibnamefont {Bhosale}},
  \bibinfo {author} {\bibfnamefont {M.}~\bibnamefont {Gazzola}}, \ and\
  \bibinfo {author} {\bibfnamefont {H.}~\bibnamefont {King}},\ }\bibfield
  {title} {\enquote {\bibinfo {title} {Mechanics of randomly packed filaments?
  the``bird nest'' as meta-material},}\ }\href@noop {} {\bibfield  {journal}
  {\bibinfo  {journal} {J. Appl. Phys.}\ }\textbf {\bibinfo {volume} {127}},\
  \bibinfo {pages} {050902} (\bibinfo {year} {2020})}\BibitemShut {NoStop}%
\bibitem [{\citenamefont {Broedersz}\ and\ \citenamefont
  {MacKintosh}(2014)}]{Broedersz:2014aa}%
  \BibitemOpen
  \bibfield  {author} {\bibinfo {author} {\bibfnamefont {C.~P.}\ \bibnamefont
  {Broedersz}}\ and\ \bibinfo {author} {\bibfnamefont {F.~C.}\ \bibnamefont
  {MacKintosh}},\ }\bibfield  {title} {\enquote {\bibinfo {title} {Modeling
  semiflexible polymer networks},}\ }\href {\doibase 10.1103/RevModPhys.86.995}
  {\bibfield  {journal} {\bibinfo  {journal} {Rev. Mod. Phys.}\ }\textbf
  {\bibinfo {volume} {86}},\ \bibinfo {pages} {995--1036} (\bibinfo {year}
  {2014})}\BibitemShut {NoStop}%
\bibitem [{\citenamefont {Philipse}(1996)}]{philipse96}%
  \BibitemOpen
  \bibfield  {author} {\bibinfo {author} {\bibfnamefont {A.~P.}\ \bibnamefont
  {Philipse}},\ }\bibfield  {title} {\enquote {\bibinfo {title} {The random
  contact equation and its implications for (colloidal) rods in packings,
  suspensions, and anisotropic powders},}\ }\href@noop {} {\bibfield  {journal}
  {\bibinfo  {journal} {Langmuir}\ }\textbf {\bibinfo {volume} {12}},\ \bibinfo
  {pages} {1127} (\bibinfo {year} {1996})}\BibitemShut {NoStop}%
\bibitem [{\citenamefont {Williams}\ and\ \citenamefont
  {Philipse}(2003)}]{williams03}%
  \BibitemOpen
  \bibfield  {author} {\bibinfo {author} {\bibfnamefont {S.~R.}\ \bibnamefont
  {Williams}}\ and\ \bibinfo {author} {\bibfnamefont {A.~P.}\ \bibnamefont
  {Philipse}},\ }\bibfield  {title} {\enquote {\bibinfo {title} {Random
  packings of spheres and spherocylinders simulated by mechanical
  contraction},}\ }\href@noop {} {\bibfield  {journal} {\bibinfo  {journal}
  {Phys. Rev. E}\ }\textbf {\bibinfo {volume} {67}},\ \bibinfo {pages} {051301}
  (\bibinfo {year} {2003})}\BibitemShut {NoStop}%
\bibitem [{\citenamefont {Desmond}\ and\ \citenamefont
  {Franklin}(2006)}]{desmond06}%
  \BibitemOpen
  \bibfield  {author} {\bibinfo {author} {\bibfnamefont {K.}~\bibnamefont
  {Desmond}}\ and\ \bibinfo {author} {\bibfnamefont {S.~V.}\ \bibnamefont
  {Franklin}},\ }\bibfield  {title} {\enquote {\bibinfo {title} {Jamming of
  three-dimensional prolate granular materials},}\ }\href@noop {} {\bibfield
  {journal} {\bibinfo  {journal} {Phys. Rev. E}\ }\textbf {\bibinfo {volume}
  {73}},\ \bibinfo {pages} {031306} (\bibinfo {year} {2006})}\BibitemShut
  {NoStop}%
\bibitem [{\citenamefont {Marschall}\ and\ \citenamefont
  {Teitel}(2018)}]{marschall18}%
  \BibitemOpen
  \bibfield  {author} {\bibinfo {author} {\bibfnamefont {T.}~\bibnamefont
  {Marschall}}\ and\ \bibinfo {author} {\bibfnamefont {S.}~\bibnamefont
  {Teitel}},\ }\bibfield  {title} {\enquote {\bibinfo {title}
  {Compression-driven jamming of athermal frictionless spherocylinders in two
  dimensions},}\ }\href@noop {} {\bibfield  {journal} {\bibinfo  {journal}
  {Phys. Rev. E}\ }\textbf {\bibinfo {volume} {97}},\ \bibinfo {pages} {012905}
  (\bibinfo {year} {2018})}\BibitemShut {NoStop}%
\bibitem [{\citenamefont {Langston}\ \emph {et~al.}(2015)\citenamefont
  {Langston}, \citenamefont {Kennedy},\ and\ \citenamefont
  {Constantin}}]{langston15}%
  \BibitemOpen
  \bibfield  {author} {\bibinfo {author} {\bibfnamefont {P.}~\bibnamefont
  {Langston}}, \bibinfo {author} {\bibfnamefont {A.~R.}\ \bibnamefont
  {Kennedy}}, \ and\ \bibinfo {author} {\bibfnamefont {H.}~\bibnamefont
  {Constantin}},\ }\bibfield  {title} {\enquote {\bibinfo {title} {Discrete
  element modelling of flexible fibre packing},}\ }\href@noop {} {\bibfield
  {journal} {\bibinfo  {journal} {Comp. Mat. Sci.}\ }\textbf {\bibinfo {volume}
  {96}},\ \bibinfo {pages} {108} (\bibinfo {year} {2015})}\BibitemShut
  {NoStop}%
\bibitem [{\citenamefont {Hoy}(2017)}]{hoy17}%
  \BibitemOpen
  \bibfield  {author} {\bibinfo {author} {\bibfnamefont {R.~S.}\ \bibnamefont
  {Hoy}},\ }\bibfield  {title} {\enquote {\bibinfo {title} {Jamming of
  semiflexible polymers},}\ }\href@noop {} {\bibfield  {journal} {\bibinfo
  {journal} {Phys. Rev. Lett.}\ }\textbf {\bibinfo {volume} {118}},\ \bibinfo
  {pages} {068002} (\bibinfo {year} {2017})}\BibitemShut {NoStop}%
\bibitem [{\citenamefont {Griffith}\ and\ \citenamefont
  {Hoy}(2019)}]{griffith19}%
  \BibitemOpen
  \bibfield  {author} {\bibinfo {author} {\bibfnamefont {A.~D.}\ \bibnamefont
  {Griffith}}\ and\ \bibinfo {author} {\bibfnamefont {R.~S.}\ \bibnamefont
  {Hoy}},\ }\bibfield  {title} {\enquote {\bibinfo {title} {Densest versus
  jammed packings of bent-core trimers},}\ }\href@noop {} {\bibfield  {journal}
  {\bibinfo  {journal} {Phys. Rev. E}\ }\textbf {\bibinfo {volume} {100}},\
  \bibinfo {pages} {022903} (\bibinfo {year} {2019})}\BibitemShut {NoStop}%
\bibitem [{\citenamefont {Karayiannis}\ and\ \citenamefont
  {Laso}(2008)}]{karayiannis08}%
  \BibitemOpen
  \bibfield  {author} {\bibinfo {author} {\bibfnamefont {N.~C.}\ \bibnamefont
  {Karayiannis}}\ and\ \bibinfo {author} {\bibfnamefont {M.}~\bibnamefont
  {Laso}},\ }\bibfield  {title} {\enquote {\bibinfo {title} {Dense and nearly
  jammed random packings of freely jointed chains of tangent hard spheres},}\
  }\href@noop {} {\bibfield  {journal} {\bibinfo  {journal} {Phys. Rev. Lett.}\
  }\textbf {\bibinfo {volume} {100}},\ \bibinfo {pages} {050602} (\bibinfo
  {year} {2008})}\BibitemShut {NoStop}%
\bibitem [{\citenamefont {Karayiannis}\ \emph
  {et~al.}(2009{\natexlab{a}})\citenamefont {Karayiannis}, \citenamefont
  {Foteinopoulou},\ and\ \citenamefont {Laso}}]{karayiannis09b}%
  \BibitemOpen
  \bibfield  {author} {\bibinfo {author} {\bibfnamefont {N.~C.}\ \bibnamefont
  {Karayiannis}}, \bibinfo {author} {\bibfnamefont {K.}~\bibnamefont
  {Foteinopoulou}}, \ and\ \bibinfo {author} {\bibfnamefont {M.}~\bibnamefont
  {Laso}},\ }\bibfield  {title} {\enquote {\bibinfo {title} {The structure of
  random packings of freely jointed chains of tangent hard spheres},}\
  }\href@noop {} {\bibfield  {journal} {\bibinfo  {journal} {J. Chem. Phys.}\
  }\textbf {\bibinfo {volume} {130}},\ \bibinfo {pages} {164908} (\bibinfo
  {year} {2009}{\natexlab{a}})}\BibitemShut {NoStop}%
\bibitem [{\citenamefont {Karayiannis}\ \emph
  {et~al.}(2009{\natexlab{b}})\citenamefont {Karayiannis}, \citenamefont
  {Foteinopoulou},\ and\ \citenamefont {Laso}}]{karayiannis09c}%
  \BibitemOpen
  \bibfield  {author} {\bibinfo {author} {\bibfnamefont {N.~C.}\ \bibnamefont
  {Karayiannis}}, \bibinfo {author} {\bibfnamefont {K.}~\bibnamefont
  {Foteinopoulou}}, \ and\ \bibinfo {author} {\bibfnamefont {M.}~\bibnamefont
  {Laso}},\ }\bibfield  {title} {\enquote {\bibinfo {title} {Contact network in
  nearly jammed disordered packings of hard-sphere chains},}\ }\href@noop {}
  {\bibfield  {journal} {\bibinfo  {journal} {Phys. Rev. E}\ }\textbf {\bibinfo
  {volume} {80}},\ \bibinfo {pages} {011307} (\bibinfo {year}
  {2009}{\natexlab{b}})}\BibitemShut {NoStop}%
\bibitem [{\citenamefont {Lopatina}\ \emph {et~al.}(2011)\citenamefont
  {Lopatina}, \citenamefont {{Olson Reichhardt}},\ and\ \citenamefont
  {Reichhardt}}]{reichhardt11}%
  \BibitemOpen
  \bibfield  {author} {\bibinfo {author} {\bibfnamefont {L.~M.}\ \bibnamefont
  {Lopatina}}, \bibinfo {author} {\bibfnamefont {C.~J.}\ \bibnamefont {{Olson
  Reichhardt}}}, \ and\ \bibinfo {author} {\bibfnamefont {C.}~\bibnamefont
  {Reichhardt}},\ }\bibfield  {title} {\enquote {\bibinfo {title} {Jamming in
  granular polymers},}\ }\href@noop {} {\bibfield  {journal} {\bibinfo
  {journal} {Phys. Rev. E}\ }\textbf {\bibinfo {volume} {84}},\ \bibinfo
  {pages} {011303} (\bibinfo {year} {2011})}\BibitemShut {NoStop}%
\bibitem [{\citenamefont {Foteinopoulou}\ \emph {et~al.}(2008)\citenamefont
  {Foteinopoulou}, \citenamefont {Karayiannis}, \citenamefont {Laso},
  \citenamefont {Kr{\"o}ger},\ and\ \citenamefont
  {Mansfield}}]{foteinopoulou08}%
  \BibitemOpen
  \bibfield  {author} {\bibinfo {author} {\bibfnamefont {K.}~\bibnamefont
  {Foteinopoulou}}, \bibinfo {author} {\bibfnamefont {N.~C.}\ \bibnamefont
  {Karayiannis}}, \bibinfo {author} {\bibfnamefont {M.}~\bibnamefont {Laso}},
  \bibinfo {author} {\bibfnamefont {M.}~\bibnamefont {Kr{\"o}ger}}, \ and\
  \bibinfo {author} {\bibfnamefont {M.~L.}\ \bibnamefont {Mansfield}},\
  }\bibfield  {title} {\enquote {\bibinfo {title} {Universal scaling,
  entanglements, and knots of model chain molecules},}\ }\href@noop {}
  {\bibfield  {journal} {\bibinfo  {journal} {Phys. Rev. Lett.}\ }\textbf
  {\bibinfo {volume} {101}},\ \bibinfo {pages} {265702} (\bibinfo {year}
  {2008})}\BibitemShut {NoStop}%
\bibitem [{\citenamefont {Rodney}\ \emph {et~al.}(2005)\citenamefont {Rodney},
  \citenamefont {Fivel},\ and\ \citenamefont {Dendievel}}]{rodney05}%
  \BibitemOpen
  \bibfield  {author} {\bibinfo {author} {\bibfnamefont {D.}~\bibnamefont
  {Rodney}}, \bibinfo {author} {\bibfnamefont {M.}~\bibnamefont {Fivel}}, \
  and\ \bibinfo {author} {\bibfnamefont {R.}~\bibnamefont {Dendievel}},\
  }\bibfield  {title} {\enquote {\bibinfo {title} {Discrete modeling of the
  mechanics of entangled materials},}\ }\href@noop {} {\bibfield  {journal}
  {\bibinfo  {journal} {Phys. Rev. Lett.}\ }\textbf {\bibinfo {volume} {95}},\
  \bibinfo {pages} {108004} (\bibinfo {year} {2005})}\BibitemShut {NoStop}%
\bibitem [{\citenamefont {Broedersz}\ \emph {et~al.}(2011)\citenamefont
  {Broedersz}, \citenamefont {Mao}, \citenamefont {Lubensky},\ and\
  \citenamefont {Mackintosh}}]{broedersz11}%
  \BibitemOpen
  \bibfield  {author} {\bibinfo {author} {\bibfnamefont {C.~P.}\ \bibnamefont
  {Broedersz}}, \bibinfo {author} {\bibfnamefont {X.}~\bibnamefont {Mao}},
  \bibinfo {author} {\bibfnamefont {T.~C.}\ \bibnamefont {Lubensky}}, \ and\
  \bibinfo {author} {\bibfnamefont {F.~C.}\ \bibnamefont {Mackintosh}},\
  }\bibfield  {title} {\enquote {\bibinfo {title} {Criticality and isostaticity
  in fibre networks},}\ }\href@noop {} {\bibfield  {journal} {\bibinfo
  {journal} {Nature Phys.}\ }\textbf {\bibinfo {volume} {7}},\ \bibinfo {pages}
  {983} (\bibinfo {year} {2011})}\BibitemShut {NoStop}%
\bibitem [{\citenamefont {Barbier}\ \emph {et~al.}(2009)\citenamefont
  {Barbier}, \citenamefont {Dendievel},\ and\ \citenamefont
  {Rodney}}]{barbier09}%
  \BibitemOpen
  \bibfield  {author} {\bibinfo {author} {\bibfnamefont {C.}~\bibnamefont
  {Barbier}}, \bibinfo {author} {\bibfnamefont {R.}~\bibnamefont {Dendievel}},
  \ and\ \bibinfo {author} {\bibfnamefont {D.}~\bibnamefont {Rodney}},\
  }\bibfield  {title} {\enquote {\bibinfo {title} {Role of friction in the
  mechanics of nonbonded fibrous materials},}\ }\href@noop {} {\bibfield
  {journal} {\bibinfo  {journal} {Phys. Rev. E}\ }\textbf {\bibinfo {volume}
  {80}},\ \bibinfo {pages} {016115} (\bibinfo {year} {2009})}\BibitemShut
  {NoStop}%
\bibitem [{\citenamefont {Subramanian}\ and\ \citenamefont
  {Picu}(2011)}]{subramanian11}%
  \BibitemOpen
  \bibfield  {author} {\bibinfo {author} {\bibfnamefont {G.}~\bibnamefont
  {Subramanian}}\ and\ \bibinfo {author} {\bibfnamefont {C.~R.}\ \bibnamefont
  {Picu}},\ }\bibfield  {title} {\enquote {\bibinfo {title} {Mechanics of
  three-dimensional, nonbonded random fiber networks},}\ }\href@noop {}
  {\bibfield  {journal} {\bibinfo  {journal} {Phys. Rev. E}\ }\textbf {\bibinfo
  {volume} {83}},\ \bibinfo {pages} {056120} (\bibinfo {year}
  {2011})}\BibitemShut {NoStop}%
\bibitem [{\citenamefont {Picu}\ and\ \citenamefont
  {Subramanian}(2011)}]{picu11b}%
  \BibitemOpen
  \bibfield  {author} {\bibinfo {author} {\bibfnamefont {R.~C.}\ \bibnamefont
  {Picu}}\ and\ \bibinfo {author} {\bibfnamefont {G.}~\bibnamefont
  {Subramanian}},\ }\bibfield  {title} {\enquote {\bibinfo {title} {Correlated
  heterogeneous deformation of entangled fiber networks},}\ }\href@noop {}
  {\bibfield  {journal} {\bibinfo  {journal} {Phys. Rev. E}\ }\textbf {\bibinfo
  {volume} {84}},\ \bibinfo {pages} {031904} (\bibinfo {year}
  {2011})}\BibitemShut {NoStop}%
\bibitem [{\citenamefont {Torquato}\ \emph {et~al.}(2000)\citenamefont
  {Torquato}, \citenamefont {Truskett},\ and\ \citenamefont
  {Debenedetti}}]{torquato00}%
  \BibitemOpen
  \bibfield  {author} {\bibinfo {author} {\bibfnamefont {S.}~\bibnamefont
  {Torquato}}, \bibinfo {author} {\bibfnamefont {T.~M.}\ \bibnamefont
  {Truskett}}, \ and\ \bibinfo {author} {\bibfnamefont {P.~G.}\ \bibnamefont
  {Debenedetti}},\ }\bibfield  {title} {\enquote {\bibinfo {title} {Is random
  close packing of spheres well defined?}}\ }\href@noop {} {\bibfield
  {journal} {\bibinfo  {journal} {Phys. Rev. Lett.}\ }\textbf {\bibinfo
  {volume} {84}},\ \bibinfo {pages} {2064} (\bibinfo {year}
  {2000})}\BibitemShut {NoStop}%
\bibitem [{\citenamefont {Chaudhuri}\ \emph {et~al.}(2010)\citenamefont
  {Chaudhuri}, \citenamefont {Berthier},\ and\ \citenamefont
  {Sastry}}]{chaudhuri10}%
  \BibitemOpen
  \bibfield  {author} {\bibinfo {author} {\bibfnamefont {P.}~\bibnamefont
  {Chaudhuri}}, \bibinfo {author} {\bibfnamefont {L.}~\bibnamefont {Berthier}},
  \ and\ \bibinfo {author} {\bibfnamefont {S.}~\bibnamefont {Sastry}},\
  }\bibfield  {title} {\enquote {\bibinfo {title} {Jamming transitions in
  amorphous packings of frictionless spheres occur over a continuous range of
  volume fractions},}\ }\href@noop {} {\bibfield  {journal} {\bibinfo
  {journal} {Phys. Rev. Lett.}\ }\textbf {\bibinfo {volume} {104}},\ \bibinfo
  {pages} {165701} (\bibinfo {year} {2010})}\BibitemShut {NoStop}%
\bibitem [{\citenamefont {Plimpton}(1995)}]{plimpton95}%
  \BibitemOpen
  \bibfield  {author} {\bibinfo {author} {\bibfnamefont {S.}~\bibnamefont
  {Plimpton}},\ }\bibfield  {title} {\enquote {\bibinfo {title} {Fast parallel
  algorithms for short-range molecular-dynamics},}\ }\href@noop {} {\bibfield
  {journal} {\bibinfo  {journal} {J. Comp. Phys.}\ }\textbf {\bibinfo {volume}
  {117}},\ \bibinfo {pages} {1} (\bibinfo {year} {1995})}\BibitemShut {NoStop}%
\bibitem [{\citenamefont {Polak}\ and\ \citenamefont
  {Ribi{\'e}re}(1969)}]{polak69}%
  \BibitemOpen
  \bibfield  {author} {\bibinfo {author} {\bibfnamefont {E.}~\bibnamefont
  {Polak}}\ and\ \bibinfo {author} {\bibfnamefont {G.}~\bibnamefont
  {Ribi{\'e}re}},\ }\bibfield  {title} {\enquote {\bibinfo {title} {Note on
  convergence of conjugate direction methods},}\ }\href@noop {} {\bibfield
  {journal} {\bibinfo  {journal} {Rev. Fr. Inf. Rech. Op.}\ }\textbf {\bibinfo
  {volume} {16}},\ \bibinfo {pages} {35} (\bibinfo {year} {1969})}\BibitemShut
  {NoStop}%
\bibitem [{\citenamefont {Grippo}\ \emph {et~al.}(1989)\citenamefont {Grippo},
  \citenamefont {Lampariello},\ and\ \citenamefont {Lucidi}}]{grippo89}%
  \BibitemOpen
  \bibfield  {author} {\bibinfo {author} {\bibfnamefont {L.}~\bibnamefont
  {Grippo}}, \bibinfo {author} {\bibfnamefont {F.}~\bibnamefont {Lampariello}},
  \ and\ \bibinfo {author} {\bibfnamefont {S.}~\bibnamefont {Lucidi}},\
  }\bibfield  {title} {\enquote {\bibinfo {title} {A truncated newton method
  with nonmonotone line search for unconstrained optimization},}\ }\href@noop
  {} {\bibfield  {journal} {\bibinfo  {journal} {J. Opt. Theor. App.}\ }\textbf
  {\bibinfo {volume} {60}},\ \bibinfo {pages} {401} (\bibinfo {year}
  {1989})}\BibitemShut {NoStop}%
\bibitem [{\citenamefont {Bitzek}\ \emph {et~al.}(2006)\citenamefont {Bitzek},
  \citenamefont {Koskinen}, \citenamefont {amd M.~Moseler},\ and\ \citenamefont
  {Gumbsch}}]{bitzek06}%
  \BibitemOpen
  \bibfield  {author} {\bibinfo {author} {\bibfnamefont {E.}~\bibnamefont
  {Bitzek}}, \bibinfo {author} {\bibfnamefont {P.}~\bibnamefont {Koskinen}},
  \bibinfo {author} {\bibfnamefont {F.~G{\"a}hler}\ \bibnamefont {amd
  M.~Moseler}}, \ and\ \bibinfo {author} {\bibfnamefont {P.}~\bibnamefont
  {Gumbsch}},\ }\bibfield  {title} {\enquote {\bibinfo {title} {Structural
  relaxation made simple},}\ }\href@noop {} {\bibfield  {journal} {\bibinfo
  {journal} {Phys. Rev. Lett.}\ }\textbf {\bibinfo {volume} {97}},\ \bibinfo
  {pages} {170201} (\bibinfo {year} {2006})}\BibitemShut {NoStop}%
\bibitem [{\citenamefont {van Hecke}(2009)}]{vanHecke09}%
  \BibitemOpen
  \bibfield  {author} {\bibinfo {author} {\bibfnamefont {M.}~\bibnamefont {van
  Hecke}},\ }\bibfield  {title} {\enquote {\bibinfo {title} {Jamming of soft
  particles: geometry, mechanics, scaling and isostaticity},}\ }\href@noop {}
  {\bibfield  {journal} {\bibinfo  {journal} {J. Phys. Cond. Matt.}\ }\textbf
  {\bibinfo {volume} {22}},\ \bibinfo {pages} {033101} (\bibinfo {year}
  {2009})}\BibitemShut {NoStop}%
\bibitem [{\citenamefont {O'Hern}\ \emph {et~al.}(2003)\citenamefont {O'Hern},
  \citenamefont {Silbert}, \citenamefont {Liu},\ and\ \citenamefont
  {Nagel}}]{ohern03}%
  \BibitemOpen
  \bibfield  {author} {\bibinfo {author} {\bibfnamefont {C.~S.}\ \bibnamefont
  {O'Hern}}, \bibinfo {author} {\bibfnamefont {L.~E.}\ \bibnamefont {Silbert}},
  \bibinfo {author} {\bibfnamefont {A.~J.}\ \bibnamefont {Liu}}, \ and\
  \bibinfo {author} {\bibfnamefont {S.~R.}\ \bibnamefont {Nagel}},\ }\bibfield
  {title} {\enquote {\bibinfo {title} {Jamming at zero temperature and zero
  applied stress: The epitome of disorder},}\ }\href@noop {} {\bibfield
  {journal} {\bibinfo  {journal} {Phys. Rev. E}\ }\textbf {\bibinfo {volume}
  {68}},\ \bibinfo {pages} {011306} (\bibinfo {year} {2003})}\BibitemShut
  {NoStop}%
\bibitem [{\citenamefont {Donev}\ \emph {et~al.}(2004)\citenamefont {Donev},
  \citenamefont {Stillinger}, \citenamefont {Chaikin},\ and\ \citenamefont
  {Torquato}}]{donev04}%
  \BibitemOpen
  \bibfield  {author} {\bibinfo {author} {\bibfnamefont {A}~\bibnamefont
  {Donev}}, \bibinfo {author} {\bibfnamefont {FH}~\bibnamefont {Stillinger}},
  \bibinfo {author} {\bibfnamefont {PM}~\bibnamefont {Chaikin}}, \ and\
  \bibinfo {author} {\bibfnamefont {S}~\bibnamefont {Torquato}},\ }\bibfield
  {title} {\enquote {\bibinfo {title} {Unusually dense crystal packings of
  ellipsoids},}\ }\href@noop {} {\bibfield  {journal} {\bibinfo  {journal}
  {Phys. Rev. Lett.}\ }\textbf {\bibinfo {volume} {92}},\ \bibinfo {pages}
  {255506} (\bibinfo {year} {2004})}\BibitemShut {NoStop}%
\bibitem [{\citenamefont {Donev}\ \emph {et~al.}(2007)\citenamefont {Donev},
  \citenamefont {Connelly}, \citenamefont {Stillinger},\ and\ \citenamefont
  {Torquato}}]{donev07}%
  \BibitemOpen
  \bibfield  {author} {\bibinfo {author} {\bibfnamefont {A.}~\bibnamefont
  {Donev}}, \bibinfo {author} {\bibfnamefont {R.}~\bibnamefont {Connelly}},
  \bibinfo {author} {\bibfnamefont {F.~H.}\ \bibnamefont {Stillinger}}, \ and\
  \bibinfo {author} {\bibfnamefont {S.}~\bibnamefont {Torquato}},\ }\bibfield
  {title} {\enquote {\bibinfo {title} {Underconstrained jammed packings of
  nonspherical hard particles: Ellipses and ellipsoids},}\ }\href@noop {}
  {\bibfield  {journal} {\bibinfo  {journal} {Phys. Rev. E}\ }\textbf {\bibinfo
  {volume} {75}},\ \bibinfo {pages} {051304} (\bibinfo {year}
  {2007})}\BibitemShut {NoStop}%
\bibitem [{\citenamefont {Hoover}\ \emph {et~al.}(1980)\citenamefont {Hoover},
  \citenamefont {Evans}, \citenamefont {Hickman}, \citenamefont {Ladd},
  \citenamefont {Ashurst},\ and\ \citenamefont {Moran}}]{hoover80}%
  \BibitemOpen
  \bibfield  {author} {\bibinfo {author} {\bibfnamefont {W.~G.}\ \bibnamefont
  {Hoover}}, \bibinfo {author} {\bibfnamefont {D.~J.}\ \bibnamefont {Evans}},
  \bibinfo {author} {\bibfnamefont {R.~B.}\ \bibnamefont {Hickman}}, \bibinfo
  {author} {\bibfnamefont {A.~J.~C.}\ \bibnamefont {Ladd}}, \bibinfo {author}
  {\bibfnamefont {W.~T.}\ \bibnamefont {Ashurst}}, \ and\ \bibinfo {author}
  {\bibfnamefont {B.}~\bibnamefont {Moran}},\ }\bibfield  {title} {\enquote
  {\bibinfo {title} {Lennard-jones triple-point bulk and shear viscosities -
  green-kubo theory, hamiltonian-mechanics, and non-equilibrium
  molecular-dynamics},}\ }\href@noop {} {\bibfield  {journal} {\bibinfo
  {journal} {Phys. Rev. A}\ }\textbf {\bibinfo {volume} {22}},\ \bibinfo
  {pages} {1690} (\bibinfo {year} {1980})}\BibitemShut {NoStop}%
\bibitem [{\citenamefont {Jiao}\ \emph {et~al.}(2010)\citenamefont {Jiao},
  \citenamefont {Stillinger},\ and\ \citenamefont {Torquato}}]{jiao10}%
  \BibitemOpen
  \bibfield  {author} {\bibinfo {author} {\bibfnamefont {Y.}~\bibnamefont
  {Jiao}}, \bibinfo {author} {\bibfnamefont {F.~H.}\ \bibnamefont
  {Stillinger}}, \ and\ \bibinfo {author} {\bibfnamefont {S.}~\bibnamefont
  {Torquato}},\ }\bibfield  {title} {\enquote {\bibinfo {title} {Distinctive
  features arising in maximally random jammed packings of superballs},}\
  }\href@noop {} {\bibfield  {journal} {\bibinfo  {journal} {Phys. Rev. E}\
  }\textbf {\bibinfo {volume} {81}},\ \bibinfo {pages} {041304} (\bibinfo
  {year} {2010})}\BibitemShut {NoStop}%
\bibitem [{Note1()}]{Note1}%
  \BibitemOpen
  \bibinfo {note} {The frictional and frictionless isostaticity criteria, $Z_J
  = 2n_{dof}/N$ and $Z_J = n_{dof}/N + 1$, are almost identical for our systems
  because $n_{dof}/N = 1 + 3/N \simeq 1$.}\BibitemShut {Stop}%
\bibitem [{\citenamefont {Schreck}\ \emph {et~al.}(2009)\citenamefont
  {Schreck}, \citenamefont {Xu},\ and\ \citenamefont {O'Hern}}]{schreck09}%
  \BibitemOpen
  \bibfield  {author} {\bibinfo {author} {\bibfnamefont {C.~F.}\ \bibnamefont
  {Schreck}}, \bibinfo {author} {\bibfnamefont {N.}~\bibnamefont {Xu}}, \ and\
  \bibinfo {author} {\bibfnamefont {C.~S.}\ \bibnamefont {O'Hern}},\ }\bibfield
   {title} {\enquote {\bibinfo {title} {A comparison of jamming behavior in
  systems composed of dimer- and ellipse-shaped particles},}\ }\href@noop {}
  {\bibfield  {journal} {\bibinfo  {journal} {Soft Matt.}\ }\textbf {\bibinfo
  {volume} {6}},\ \bibinfo {pages} {2960} (\bibinfo {year} {2009})}\BibitemShut
  {NoStop}%
\bibitem [{\citenamefont {Ludewig}\ and\ \citenamefont
  {Vandewalle}(2012)}]{ludewig12}%
  \BibitemOpen
  \bibfield  {author} {\bibinfo {author} {\bibfnamefont {F.}~\bibnamefont
  {Ludewig}}\ and\ \bibinfo {author} {\bibfnamefont {N.}~\bibnamefont
  {Vandewalle}},\ }\bibfield  {title} {\enquote {\bibinfo {title} {Strong
  interlocking of nonconvex particles in random packings},}\ }\href@noop {}
  {\bibfield  {journal} {\bibinfo  {journal} {Phys. Rev. E}\ }\textbf {\bibinfo
  {volume} {85}},\ \bibinfo {pages} {051307} (\bibinfo {year}
  {2012})}\BibitemShut {NoStop}%
\bibitem [{\citenamefont {Thorpe}(1985)}]{thorpe85}%
  \BibitemOpen
  \bibfield  {author} {\bibinfo {author} {\bibfnamefont {M.~F.}\ \bibnamefont
  {Thorpe}},\ }\bibfield  {title} {\enquote {\bibinfo {title} {Rigidity
  percolation in glassy structures},}\ }\href@noop {} {\bibfield  {journal}
  {\bibinfo  {journal} {J. Non-cryst. Solids}\ }\textbf {\bibinfo {volume}
  {76}},\ \bibinfo {pages} {109} (\bibinfo {year} {1985})}\BibitemShut
  {NoStop}%
\bibitem [{\citenamefont {Bekker}\ and\ \citenamefont
  {Ahlstr{\"o}m}(1994)}]{bekker94}%
  \BibitemOpen
  \bibfield  {author} {\bibinfo {author} {\bibfnamefont {H.}~\bibnamefont
  {Bekker}}\ and\ \bibinfo {author} {\bibfnamefont {P.}~\bibnamefont
  {Ahlstr{\"o}m}},\ }\bibfield  {title} {\enquote {\bibinfo {title} {The virial
  of angle dependent potentials in molecular dynamics simulations},}\
  }\href@noop {} {\bibfield  {journal} {\bibinfo  {journal} {Mol. Sim.}\
  }\textbf {\bibinfo {volume} {13}},\ \bibinfo {pages} {367} (\bibinfo {year}
  {1994})}\BibitemShut {NoStop}%
\bibitem [{\citenamefont {{de Gennes}}(1974)}]{degennes74}%
  \BibitemOpen
  \bibfield  {author} {\bibinfo {author} {\bibfnamefont {P.~G.}\ \bibnamefont
  {{de Gennes}}},\ }\bibfield  {title} {\enquote {\bibinfo {title} {Remarks on
  entanglements and rubber elasticity},}\ }\href@noop {} {\bibfield  {journal}
  {\bibinfo  {journal} {J. Phys. Lett}\ }\textbf {\bibinfo {volume} {35}},\
  \bibinfo {pages} {L133} (\bibinfo {year} {1974})}\BibitemShut {NoStop}%
\bibitem [{\citenamefont {Colby}\ \emph {et~al.}(1992)\citenamefont {Colby},
  \citenamefont {Rubinstein},\ and\ \citenamefont {Viovy}}]{colby92}%
  \BibitemOpen
  \bibfield  {author} {\bibinfo {author} {\bibfnamefont {R.~H.}\ \bibnamefont
  {Colby}}, \bibinfo {author} {\bibfnamefont {M.}~\bibnamefont {Rubinstein}}, \
  and\ \bibinfo {author} {\bibfnamefont {J.~L.}\ \bibnamefont {Viovy}},\
  }\bibfield  {title} {\enquote {\bibinfo {title} {Chain entanglement in
  polymer melts and solutions},}\ }\href@noop {} {\bibfield  {journal}
  {\bibinfo  {journal} {Macromolecules}\ }\textbf {\bibinfo {volume} {25}},\
  \bibinfo {pages} {996} (\bibinfo {year} {1992})}\BibitemShut {NoStop}%
\bibitem [{Note2()}]{Note2}%
  \BibitemOpen
  \bibinfo {note} {This two-stage mechanism appears to be specific to \protect
  \textit {quasistatic} compression; FR polymers under dynamic compression
  exhibit increases in $Z_{\protect \rm chain}(\phi )$ that are gradual rather
  than sharp, and viscous ($\protect \mathaccentV {dot}05F\epsilon $-dependent)
  stresses within the liquid phase.\cite {hoy17}}\BibitemShut {NoStop}%
\bibitem [{Note3()}]{Note3}%
  \BibitemOpen
  \bibinfo {note} {Our previous study \cite {hoy17} defined flippers as
  interior monomers with less than two noncovalent contacts. We find that the
  more restrictive definition employed here clarifies the relevant
  physics.}\BibitemShut {Stop}%
\bibitem [{Note4()}]{Note4}%
  \BibitemOpen
  \bibinfo {note} {$n_\theta = 360^\circ /\theta _0$ consecutive $\Psi = 0$
  conformers form a circle; smaller numbers of consecutive $\Psi = 0$
  conformers form a planar circular arc.}\BibitemShut {Stop}%
\bibitem [{\citenamefont {Anikeenko}\ and\ \citenamefont
  {Medvedev}(2007)}]{anikeenko07}%
  \BibitemOpen
  \bibfield  {author} {\bibinfo {author} {\bibfnamefont {A/~V.}\ \bibnamefont
  {Anikeenko}}\ and\ \bibinfo {author} {\bibfnamefont {N.~N.}\ \bibnamefont
  {Medvedev}},\ }\bibfield  {title} {\enquote {\bibinfo {title}
  {Polytetrahedral nature of the dense disordered packings of hard spheres},}\
  }\href@noop {} {\bibfield  {journal} {\bibinfo  {journal} {Phys. Rev. Lett.}\
  }\textbf {\bibinfo {volume} {98}},\ \bibinfo {pages} {235504} (\bibinfo
  {year} {2007})}\BibitemShut {NoStop}%
\bibitem [{Note5()}]{Note5}%
  \BibitemOpen
  \bibinfo {note} {This collapse mechanism appears to be specific to \protect
  \textit {quasistatic} compression; large-$\theta _0$ FR polymers under
  dynamic compression do not develop sharp peaks in $P(|\Psi |)$.\cite
  {hoy17}}\BibitemShut {NoStop}%
\bibitem [{\citenamefont {O'Hern}\ \emph {et~al.}(2001)\citenamefont {O'Hern},
  \citenamefont {Langer}, \citenamefont {Liu},\ and\ \citenamefont
  {Nagel}}]{ohern01}%
  \BibitemOpen
  \bibfield  {author} {\bibinfo {author} {\bibfnamefont {C.~S.}\ \bibnamefont
  {O'Hern}}, \bibinfo {author} {\bibfnamefont {S.~A.}\ \bibnamefont {Langer}},
  \bibinfo {author} {\bibfnamefont {A.~J.}\ \bibnamefont {Liu}}, \ and\
  \bibinfo {author} {\bibfnamefont {S.~R.}\ \bibnamefont {Nagel}},\ }\bibfield
  {title} {\enquote {\bibinfo {title} {Force distributions near jamming and
  glass transitions},}\ }\href@noop {} {\bibfield  {journal} {\bibinfo
  {journal} {Phys. Rev. Lett.}\ }\textbf {\bibinfo {volume} {86}},\ \bibinfo
  {pages} {111} (\bibinfo {year} {2001})}\BibitemShut {NoStop}%
\bibitem [{\citenamefont {Rottler}\ and\ \citenamefont
  {Robbins}(2002)}]{rottler02b}%
  \BibitemOpen
  \bibfield  {author} {\bibinfo {author} {\bibfnamefont {J.}~\bibnamefont
  {Rottler}}\ and\ \bibinfo {author} {\bibfnamefont {M.~O.}\ \bibnamefont
  {Robbins}},\ }\bibfield  {title} {\enquote {\bibinfo {title} {Jamming under
  tension in polymer crazes},}\ }\href@noop {} {\bibfield  {journal} {\bibinfo
  {journal} {Phys. Rev. Lett.}\ }\textbf {\bibinfo {volume} {89}},\ \bibinfo
  {pages} {195501} (\bibinfo {year} {2002})}\BibitemShut {NoStop}%
\bibitem [{\citenamefont {Peters}\ \emph {et~al.}(2005)\citenamefont {Peters},
  \citenamefont {Muthuswamy}, \citenamefont {Wibowo},\ and\ \citenamefont
  {Tordesillas}}]{peters05}%
  \BibitemOpen
  \bibfield  {author} {\bibinfo {author} {\bibfnamefont {J.~F.}\ \bibnamefont
  {Peters}}, \bibinfo {author} {\bibfnamefont {M.}~\bibnamefont {Muthuswamy}},
  \bibinfo {author} {\bibfnamefont {J.}~\bibnamefont {Wibowo}}, \ and\ \bibinfo
  {author} {\bibfnamefont {A.}~\bibnamefont {Tordesillas}},\ }\bibfield
  {title} {\enquote {\bibinfo {title} {Characterization of force chains in
  granular material},}\ }\href@noop {} {\bibfield  {journal} {\bibinfo
  {journal} {Phys. Rev. E}\ }\textbf {\bibinfo {volume} {72}},\ \bibinfo
  {pages} {041307} (\bibinfo {year} {2005})}\BibitemShut {NoStop}%
\bibitem [{\citenamefont {Meenakshisundaram}\ \emph {et~al.}(2019)\citenamefont
  {Meenakshisundaram}, \citenamefont {Hung},\ and\ \citenamefont
  {Simmons}}]{meena19}%
  \BibitemOpen
  \bibfield  {author} {\bibinfo {author} {\bibfnamefont {V.}~\bibnamefont
  {Meenakshisundaram}}, \bibinfo {author} {\bibfnamefont {{J.-H.}}\
  \bibnamefont {Hung}}, \ and\ \bibinfo {author} {\bibfnamefont {D.~S.}\
  \bibnamefont {Simmons}},\ }\bibfield  {title} {\enquote {\bibinfo {title}
  {Design rules for glass formation from model molecules designed by a
  neural-network-biased genetic algorithm},}\ }\href@noop {} {\bibfield
  {journal} {\bibinfo  {journal} {Soft Matt.}\ }\textbf {\bibinfo {volume}
  {15}} (\bibinfo {year} {2019})}\BibitemShut {NoStop}%
\bibitem [{\citenamefont {Salcedo}\ \emph {et~al.}(2019)\citenamefont
  {Salcedo}, \citenamefont {Nguyen},\ and\ \citenamefont {Hoy}}]{salcedo19}%
  \BibitemOpen
  \bibfield  {author} {\bibinfo {author} {\bibfnamefont {E.~T.}\ \bibnamefont
  {Salcedo}}, \bibinfo {author} {\bibfnamefont {H.~T.}\ \bibnamefont {Nguyen}},
  \ and\ \bibinfo {author} {\bibfnamefont {R.~S.}\ \bibnamefont {Hoy}},\
  }\bibfield  {title} {\enquote {\bibinfo {title} {Factors influencing thermal
  solidification of bent-core trimers},}\ }\href@noop {} {\bibfield  {journal}
  {\bibinfo  {journal} {J. Chem. Phys.}\ }\textbf {\bibinfo {volume} {151}},\
  \bibinfo {pages} {134501} (\bibinfo {year} {2019})}\BibitemShut {NoStop}%
\bibitem [{\citenamefont {Zou}\ \emph {et~al.}(2009)\citenamefont {Zou},
  \citenamefont {Cheng}, \citenamefont {Rivers}, \citenamefont {Jaeger},\ and\
  \citenamefont {Nagel}}]{zou09}%
  \BibitemOpen
  \bibfield  {author} {\bibinfo {author} {\bibfnamefont {L.-N.}\ \bibnamefont
  {Zou}}, \bibinfo {author} {\bibfnamefont {X.}~\bibnamefont {Cheng}}, \bibinfo
  {author} {\bibfnamefont {M.~L.}\ \bibnamefont {Rivers}}, \bibinfo {author}
  {\bibfnamefont {H.~M.}\ \bibnamefont {Jaeger}}, \ and\ \bibinfo {author}
  {\bibfnamefont {S.~R.}\ \bibnamefont {Nagel}},\ }\bibfield  {title} {\enquote
  {\bibinfo {title} {The packing of granular polymer chains},}\ }\href@noop {}
  {\bibfield  {journal} {\bibinfo  {journal} {Science}\ }\textbf {\bibinfo
  {volume} {326}},\ \bibinfo {pages} {408--410} (\bibinfo {year}
  {2009})}\BibitemShut {NoStop}%
\bibitem [{\citenamefont {Brown}\ \emph {et~al.}(2012)\citenamefont {Brown},
  \citenamefont {Nasto}, \citenamefont {Athanassiadis},\ and\ \citenamefont
  {Jaeger}}]{brown12}%
  \BibitemOpen
  \bibfield  {author} {\bibinfo {author} {\bibfnamefont {E.}~\bibnamefont
  {Brown}}, \bibinfo {author} {\bibfnamefont {A.}~\bibnamefont {Nasto}},
  \bibinfo {author} {\bibfnamefont {A.~G.}\ \bibnamefont {Athanassiadis}}, \
  and\ \bibinfo {author} {\bibfnamefont {H.~M.}\ \bibnamefont {Jaeger}},\
  }\bibfield  {title} {\enquote {\bibinfo {title} {Strain stiffening in random
  packings of entangled granular chains},}\ }\href@noop {} {\bibfield
  {journal} {\bibinfo  {journal} {Phys. Rev. Lett.}\ }\textbf {\bibinfo
  {volume} {108}},\ \bibinfo {pages} {108302} (\bibinfo {year}
  {2012})}\BibitemShut {NoStop}%
\bibitem [{\citenamefont {Vutukuri}\ \emph {et~al.}(2012)\citenamefont
  {Vutukuri}, \citenamefont {Demir{\"o}rs}, \citenamefont {Peng}, \citenamefont
  {{van Oostrum}}, \citenamefont {Imhof},\ and\ \citenamefont {{van
  Blaaderen}}}]{vutukuri12}%
  \BibitemOpen
  \bibfield  {author} {\bibinfo {author} {\bibfnamefont {H.~R.}\ \bibnamefont
  {Vutukuri}}, \bibinfo {author} {\bibfnamefont {A.~F.}\ \bibnamefont
  {Demir{\"o}rs}}, \bibinfo {author} {\bibfnamefont {B.}~\bibnamefont {Peng}},
  \bibinfo {author} {\bibfnamefont {P.~D.~J.}\ \bibnamefont {{van Oostrum}}},
  \bibinfo {author} {\bibfnamefont {A.}~\bibnamefont {Imhof}}, \ and\ \bibinfo
  {author} {\bibfnamefont {A.}~\bibnamefont {{van Blaaderen}}},\ }\bibfield
  {title} {\enquote {\bibinfo {title} {Colloidal analogues of charged and
  uncharged polymer chains with tunable stiffness},}\ }\href@noop {} {\bibfield
   {journal} {\bibinfo  {journal} {Angew. Chem. Int. Ed. Engl.}\ }\textbf
  {\bibinfo {volume} {51}},\ \bibinfo {pages} {11249} (\bibinfo {year}
  {2012})}\BibitemShut {NoStop}%
\bibitem [{\citenamefont {Mcmullen}\ \emph {et~al.}(2018)\citenamefont
  {Mcmullen}, \citenamefont {Holmes-Cerfon}, \citenamefont {Sciortino},
  \citenamefont {Grosberg},\ and\ \citenamefont {Brujic}}]{mcmullen18}%
  \BibitemOpen
  \bibfield  {author} {\bibinfo {author} {\bibfnamefont {A.}~\bibnamefont
  {Mcmullen}}, \bibinfo {author} {\bibfnamefont {M.}~\bibnamefont
  {Holmes-Cerfon}}, \bibinfo {author} {\bibfnamefont {F.}~\bibnamefont
  {Sciortino}}, \bibinfo {author} {\bibfnamefont {A.~Y.}\ \bibnamefont
  {Grosberg}}, \ and\ \bibinfo {author} {\bibfnamefont {J.}~\bibnamefont
  {Brujic}},\ }\bibfield  {title} {\enquote {\bibinfo {title} {Freely jointed
  polymers made of droplets},}\ }\href@noop {} {\bibfield  {journal} {\bibinfo
  {journal} {Phys. Rev. Lett.}\ }\textbf {\bibinfo {volume} {121}},\ \bibinfo
  {pages} {138002} (\bibinfo {year} {2018})}\BibitemShut {NoStop}%
\bibitem [{\citenamefont {Dumont}\ \emph {et~al.}(2018)\citenamefont {Dumont},
  \citenamefont {Houze}, \citenamefont {Rambach}, \citenamefont {Salez},
  \citenamefont {Patinet},\ and\ \citenamefont {Damman}}]{dumont18}%
  \BibitemOpen
  \bibfield  {author} {\bibinfo {author} {\bibfnamefont {D.}~\bibnamefont
  {Dumont}}, \bibinfo {author} {\bibfnamefont {M.}~\bibnamefont {Houze}},
  \bibinfo {author} {\bibfnamefont {P.}~\bibnamefont {Rambach}}, \bibinfo
  {author} {\bibfnamefont {T.}~\bibnamefont {Salez}}, \bibinfo {author}
  {\bibfnamefont {S.}~\bibnamefont {Patinet}}, \ and\ \bibinfo {author}
  {\bibfnamefont {P.}~\bibnamefont {Damman}},\ }\bibfield  {title} {\enquote
  {\bibinfo {title} {Emergent strain stiffening in interlocked granular
  chains},}\ }\href@noop {} {\bibfield  {journal} {\bibinfo  {journal} {Phys.
  Rev. Lett,}\ }\textbf {\bibinfo {volume} {120}},\ \bibinfo {pages} {088001}
  (\bibinfo {year} {2018})}\BibitemShut {NoStop}%
\end{thebibliography}


\end{document}